\documentstyle [11pt,cite,epsfig,rotating]{article}
\def\q2{$Q^2$}
\title{\begin{center}  
{\sc Accurate Measurement of $F_{2}^{\rm d}/F_{2}^{\rm p}$
and $R^{\rm d}-R^{\rm p}$}
\end{center}}

\author{}
\date{}
\begin{document}

\begin{center}
\maketitle
\end{center}
\vspace{0.6cm}




\begin{center} 
THE NEW MUON COLLABORATION (NMC) \\
\vspace{0.6cm}
{\footnotesize{\sl{
Bielefeld~University$^{1+}$,
Freiburg~University$^{2+}$,
Max-Planck~Institut f\"{u}r Kernphysik, Heidelberg$^{3+}$,
Heidelberg~University$^{4+}$,
Mainz~University$^{5+}$, Mons~University$^6$,
Neuch\^{a}tel~University$^7$,
NIKHEF$^{8++}$,
Saclay DAPNIA/SPP$^{9**}$,
University~of~California, Santa~Cruz$^{10}$,
Paul~Scherrer~Institut$^{11}$,
Torino~University and INFN~Torino$^{12}$,
Uppsala~University$^{13}$,
Soltan~Institute~for~Nuclear~Studies, Warsaw$^{14*}$,
Warsaw~University$^{15*}$}}}\\
\vspace{0.6cm}
{\small{M.~Arneodo$^{12}$,
A.~Arvidson$^{13}$,
B.~Bade{\l }ek$^{13,15}$,
M.~Ballintijn$^8$,
G.~Baum$^1$,
J.~Beaufays$^8$,
I.G.~Bird$^{3,8a)}$,
P.~Bj\"{o}rkholm$^{13}$,
M.~Botje$^{11b)}$,
C.~Broggini$^{7c)}$,
W.~Br\"{u}ckner$^3$,
A.~Br\"{u}ll$^{2d)}$,
W.J.~Burger$^{11e)}$,
J.~Ciborowski$^{15}$,
R.~van~Dantzig$^8$,
A.~Dyring$^{13}$,
H.~Engelien$^2$,
M.I.~Ferrero$^{12}$,
L.~Fluri$^7$,
U.~Gaul$^3$,
T.~Granier$^{9f)}$,
M.~Grosse-Perdekamp$^{2g)}$,
D.~von~Harrach$^{3h)}$,
M.~van~der~Heijden$^8$,
C.~Heusch$^{10}$,
Q.~Ingram$^{11}$,
M.~de~Jong$^{8a)}$,
E.M.~Kabu\ss$^{3h)}$,
R.~Kaiser$^2$,
T.J.~Ketel$^8$,
F.~Klein$^{5i)}$,
S.~Kullander$^{13}$,
U.~Landgraf$^2$,
T.~Lindqvist$^{13}$,
G.K.~Mallot$^{5}$,
C.~Mariotti$^{12j)}$,
G.~van~Middelkoop$^{8}$,
A.~Milsztajn$^9$,
Y.~Mizuno$^{3k)}$,
A.~Most$^{3l)}$,
A.~M\"{u}cklich$^3$,
J.~Nassalski$^{14}$,
D.~Nowotny$^{3}$,
J.~Oberski$^8$,
A.~Pai\'{c}$^7$,
C.~Peroni$^{12}$,
B.~Povh$^{3,4}$,
K.~Prytz$^{13m)}$,
R.~Rieger$^{5}$,
K.~Rith$^{3n)}$,
K.~R\"{o}hrich$^{5o)}$,
E.~Rondio$^{14a)}$,
L.~Ropelewski$^{15a)}$,
A.~Sandacz$^{14}$,
D.~Sanders$^{p)}$,
C.~Scholz$^{3}$,
R.~Seitz$^{5q)}$,
F.~Sever$^{1,8r)}$,
T.-A.~Shibata$^{4s)}$,
M.~Siebler$^1$,
A.~Simon$^{3t)}$,
A.~Staiano$^{12}$,
M.~Szleper$^{14}$,
W.~T{\l }acza{\l }a$^{14u)}$
Y.~Tzamouranis$^{3p)}$,
M.~Virchaux$^9$,
J.L.~Vuilleumier$^7$,
T.~Walcher$^5$,
R.~Windmolders$^6$,
A.~Witzmann$^2$,
K.~Zaremba$^{14u)}$,
F.~Zetsche$^{3v)}$}} \\
\vspace{0.6cm}
\end{center}

\begin{center}
{\footnotesize\it (to be submitted to Nuclear Physics)}
\end{center}
\noindent
\begin{abstract}
Results are presented for $F_{2}^{\rm d}/F_{2}^{\rm p}$ and
$R^{\rm d}-R^{\rm p}$
from simultaneous measurements of deep inelastic muon
scattering
on hydrogen and deuterium targets, at 90, 120, 200 and 280 GeV.
The difference $R^{\rm d}-R^{\rm p}$, determined
in the range $0.002<x<0.4$ at an average $Q^2$ of 5 GeV$^2$, is
compatible with zero. The $x$ and $Q^2$ dependence of
$F_{2}^{\rm d}/F_{2}^{\rm p}$ was measured in the kinematic range
$0.001<x<0.8$ and $0.1<Q^2<145$~GeV$^2$ with small statistical and
systematic errors.
For $x>0.1$ the ratio decreases with $Q^2$.
\end{abstract}
 
{\footnotesize {-----------------------------------\\
 
For footnotes see next page.}}
\newpage

\begin{tabbing}
~~~~\=+~~~\=Supported by Bundesministerium f\"{u}r Bildung und Forschung.\\
\>      ++\>    Supported in part by FOM, Vrije Universiteit Amsterdam and NWO.\\
\>       *\>    Supported by KBN SPUB Nr 621/E - 78/SPUB/P3/209/94. \\
\>      **\>    Laboratory of CEA, Direction des Sciences de la Mati\a`ere.\\
\> \> \\
\>      a)\>    Now at CERN, 1211 Gen\a`eve 23, Switzerland. \\
\>      b)\>    Now at NIKHEF, 1009 DB Amsterdam, The Netherlands. \\
\>      c)\>    Now at University of Padova, 35131 Padova, Italy.\\
\>      d)\>    Now at MPI f\"{u}r Kernphysik, 69029 Heidelberg, Germany. \\
\>      e)\>    Now at Universit\a'e de Gen\a`eve, 1211 Gen\a`eve 4, Switzerland.\\
\>      f)\>    Now at DPTA, CEA, Bruy\a`eres-le-Chatel, France. \\
\>      g)\>    Now at Yale University, New Haven, 06511 CT, U.S.A.\\
\>      h)\>    Now at University of Mainz, 55099 Mainz, Germany. \\
\>      i)\>    Now at University of Bonn, 53115 Bonn, Germany. \\
\>      j)\>    Now at INFN-Istituto Superiore di Sanit\a`a, 00161 Roma, Italy.\\
\>      k)\>    Now at Osaka University, 567 Osaka, Japan.\\
\>      l)\>    Now at University of Michigan, Michigan, U.S.A.\\
\>      m)\>    Now at Stockholm University, 113 85 Stockholm, Sweden.\\
\>      n)\>    Now at University of Erlangen-N\"{u}rnberg, 91058 Erlangen, Germany.\\
\>      o)\>    Now at IKP2-KFA, 52428 J\"{u}lich, Germany.\\
\>      p)\>    Now at University of Houston, 77204 TX, U.S.A.\\
\>      q)\>    Now at Dresden University, 01062 Dresden, Germany.\\
\>      r)\>    Now at ESRF, 38043 Grenoble, France.\\
\>      s)\>    Now at Tokyo Institute of Technology, Tokyo, Japan. \\
\>      t)\>    Now at University of Freiburg, 79104 Freiburg, Germany. \\
\>      u)\>    Now at Warsaw University of Technology, Warsaw, Poland. \\
\>      v)\>    Now at University of Hamburg, 22761 Hamburg, Germany.\\

\end{tabbing}
\vskip 0.5 cm
\vspace{0.8cm}
 
\newpage
\section{Introduction}

In this paper we present
an accurate, high statistics measurement of
the ratio of the structure functions of the
deuteron and the proton, $F_{2}^{\rm d}/F_{2}^{\rm p}$,
and of the difference, $R^{\rm d}-R^{\rm p}$,
obtained in deep inelastic muon
scattering at  incident
energies of 90, 120, 200 and 280 GeV.
Here, $R$ is the ratio of longitudinally to transversely polarised virtual
photon absorption cross sections.
The main motivations for this measurement are as follows:
\begin{itemize}
\item
From the measured ratio
$F_{2}^{\rm d}/F_{2}^{\rm p}$, the ratio of the neutron
and proton structure functions can be extracted.
In the parton picture of the nucleon
$F_{2}^{\rm n}/F_{2}^{\rm p}$ is related to
the ratio of the down and up quark momentum distributions.
Thus, a precise measurement of $F_{2}^{\rm d}/F_{2}^{\rm p}$
puts strong constraints on the flavour
composition of the nucleon as a function of the quark momentum.
\item
Although the proton and neutron have different flavour compositions,
the $Q^2$ dependences of $F_2^{\rm p}$ and $F_2^{\rm d}$ are 
similar,
resulting in a slight
$Q^2$ dependence of $F_{2}^{\rm d}/F_{2}^{\rm p}$ which can be 
calculated in perturbative QCD.
\item
Perturbative QCD also predicts that $\Delta R =R^{\rm d}-R^{\rm p}$
is sensitive to differences in the gluon distributions of the proton
and the neutron. Thus, through a measurement of $\Delta R$ 
one can estimate such differences.
\end{itemize}
The differential cross section for one photon exchange can be written in
terms of the nucleon structure function $F_2(x,Q^2)$ and the ratio $R(x,Q^2)$
as
\begin{eqnarray}
\frac{{\rm d}^2\sigma(x,Q^2,E)}{{\rm d}x {\rm d} Q^2} &=&
\frac{4 \pi \alpha^2 }{Q^4} \cdot
\frac{F_2(x,Q^2)}{x} \cdot \nonumber \\
 && \left\{ 1 -y -\frac{Q^2}{4E^2}
+ (1 - \frac{2m^2}{Q^2}) \cdot \frac{y^2 +Q^2/E^2}
{2\left(1+R(x,Q^2)\right)} \right\}\,\,,
\label{eq:sigma}
\end{eqnarray}
where $\alpha$ is the fine structure constant,
$-Q^2$ the four-momentum transfer squared, $E$ the energy of the
incident muon and $m$ the muon mass.  The Bjorken scaling variable,
$x$, and $y$ are defined as $x=Q^2/2M\nu$
and $y=\nu/E$, where $\nu$ is the energy of the virtual
photon in the target rest frame and $M$ the proton mass.
Throughout this paper, cross sections and structure functions are always given 
per nucleon.

To extract the structure function ratio, $F_{2}^{\rm d}/F_{2}^{\rm p}$,
and the difference, $R^{\rm d}-R^{\rm p}$, from the cross section ratio,
$\sigma_{\rm d}/\sigma_{\rm p}$, measurements at different incident
energies with a large overlap in $x$ and $Q^2$ are needed. If 
$R^{\rm d}=R^{\rm p}$, the ratios $\sigma_{\rm d}/\sigma_{\rm p}$
and $F_{2}^{\rm d}/F_{2}^{\rm p}$ are equal, as is apparent from 
eq.(\ref{eq:sigma}).

In the present experiment the ratio
$\sigma_{\rm d}/\sigma_{\rm p}$ was obtained from a
simultaneous measurement on hydrogen and deuterium in a symmetric target
arrangement. This results in a cancellation of systematic errors due to
spectrometer acceptance and normalisation and allows measurements in
kinematic regions where the detector acceptance is small.

The results presented here use the full NMC proton and deuteron data;
they supersede those already published in 
refs.\,\cite{nmcrat,nmcratlong,nmcdeltar}
that were based on part of the data.
They cover a broad kinematic range $0.001<x<0.8$ and
$0.1<Q^2<145$~GeV$^2$ and have total systematic errors typically below 1\%.

\section{The experiment}

\begin{figure}[htb]
\begin{center}
\mbox{\epsfig{figure=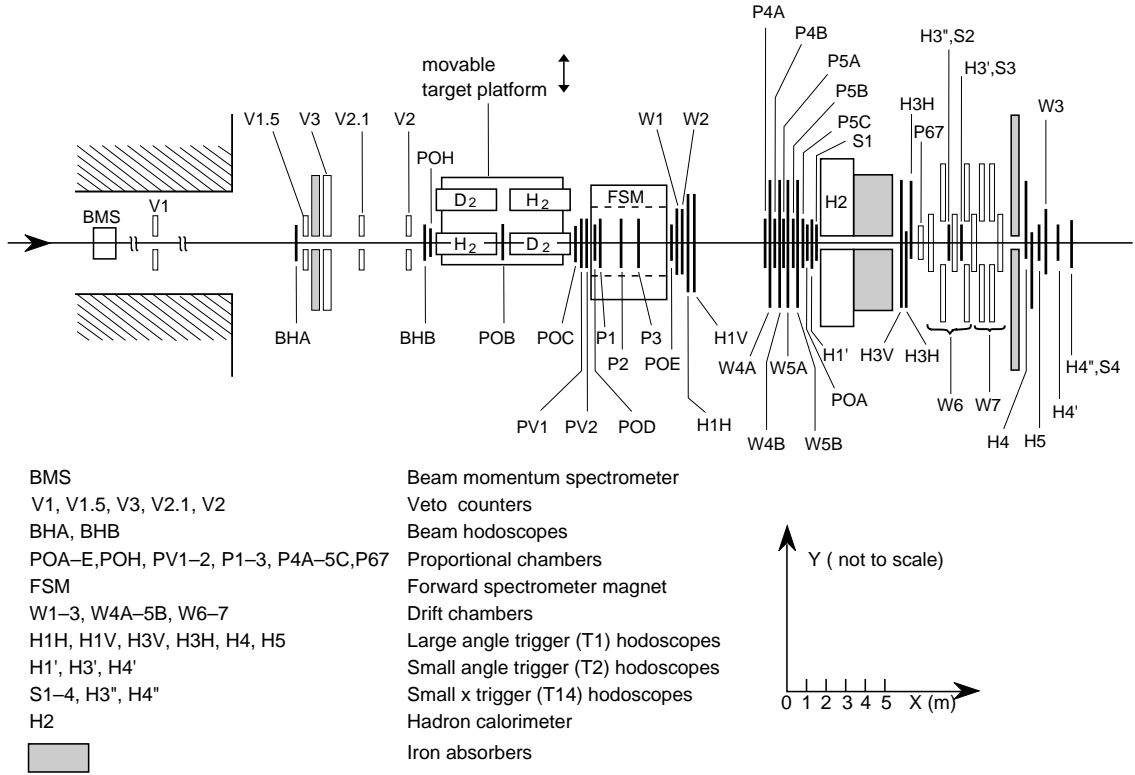,width=\textwidth}
 }
\end{center}
\caption{The NMC spectrometer for the 1989 data taking. The beam calibration
spectrometer is located downstream and not shown.}
\label{detector}
\end{figure}

The NA37 experiment was performed at the M2 muon beam line
of the CERN SPS.
The data were taken in 1986 and 1987 at nominal incident energies of
90 and 280 GeV, and in 1989 at 120, 200 and 280 GeV.
The spectrometer is described in detail in
refs.\,\cite{nmcratlong,apparatus} and the layout for
the 1989 run is shown in fig.\,\ref{detector}.

The momenta of the incoming muons were determined with a beam momentum
 spectrometer
(BMS) and their positions in two hodoscopes (BHA, BHB) upstream of the targets.
Scattered muons and produced hadrons were measured in a forward
spectrometer consisting of a dipole magnet (FSM),
proportional chambers (P), drift chambers (W) and trigger hodoscopes (H, S).
Particles passing through a 2 m thick iron absorber were identified as muons.

The measurements were performed simultaneously on hydrogen and deuterium.
The target system
contained two sets of target pairs which were alternately exposed to the
beam. In one pair the upstream target was liquid hydrogen and the downstream one
liquid deuterium, while in the other pair the order was reversed.
Frequent exchange of the two target sets (typically twice per hour) minimised
the effect of any time dependent detector response.
The targets were contained in 3 m long
mylar cells. Their thicknesses were 21.06(1) g/cm$^2$ for H$_2$ and
48.58(1) g/cm$^2$
for D$_2$ with a 3.0(2)\% HD admixture in the D$_2$.
The total amount of mylar in the beam was 0.12 g/cm$^2$ per
target, including target superinsulation.

For the 1989 data taking, upgrades were made to extend the accessible
kinematic range towards smaller scattering angles, $\theta$,
thus allowing smaller values of $x$ to be reached.
The longitudinal vertex resolution was improved by adding an extra
tracking chamber with 1 mm wire spacing (P0H) in front of the upstream target.
To detect muons scattered at small angles, an additional trigger system (T14)
was set up in addition to the two triggers (T1,T2)
described in ref.\,\cite{nmcratlong}, by
using small scintillators (S1, S2, S4) placed just
above and below the muon beam \cite{seitz}. In this trigger only the central
 part of the beam
was used to avoid triggering on divergent beam tracks.
Additional tracking chambers (P67, W3) were installed behind the iron
absorber to improve the reconstruction of small angle tracks.
Also, the performance of the small angle trigger (T2) was
improved considerably so that  a large increase in the yield at small
$x$ values was obtained in 1989.

The calibration of the scattered and incident muon momenta was done using
various methods. The forward spectrometer magnet was calibrated to an accuracy
of 0.2\% by comparing the observed J/$\psi$ and K$^0$ masses with their known
values. The beam momentum spectrometer was calibrated in dedicated
runs by remeasuring the incident muon momentum in a purpose built spectrometer
\cite{bcs}. An independent calibration of the BMS 
relative to the FSM was obtained
using silicon microstrip detectors \cite{anna}.
The two BMS calibrations
were averaged, leading to an accuracy in the incident muon momentum  of 0.2\%.

\section{Data analysis}

\subsection{Extraction of $\sigma_{\rm d}/\sigma_{\rm p}$}

The ratio of cross sections for the deuteron and the proton was obtained
from the
measured numbers of events in the four targets.
The description of the event reconstruction can be found in
refs.\,\cite{nmcratlong,apparatus,seitz}.
In any ($x$,$Q^2$) bin the number of scattered muons detected in the
spectrometer and originating
e.g. in the upstream hydrogen target is given by
\begin{equation}
N_{\rm p}^{\rm up}=\phi_1 \rho_{\rm p} \sigma_{\rm p}^{\rm incl}
A_{\rm p}^{\rm up}.
\label{eq:events}
\end{equation}
Here $\phi_1$ is the integrated beam flux illuminating the targets of the
first set, $\rho_{\rm p}$ the number of target nucleons per unit area,
$\sigma_{\rm p}^{\rm incl}$
the inclusive cross section per nucleon and $A_{\rm p}^{\rm up}$ the
 acceptance.
With equivalent expressions for the other three targets one obtains
\begin{equation}
\frac{\sigma_{\rm d}^{\rm incl}}{\sigma_{\rm
 p}^{\rm incl}}=\frac{\rho_{\rm p}}{\rho_{\rm d}}
\sqrt{\frac{N_{\rm d}^{\rm up}N_{\rm d}^{\rm down}}
{N_{\rm p}^{\rm up}N_{\rm p}^{\rm down}}}\,\,,
\label{eq:ratio}
\end{equation}
under the assumption
that $A_{\rm d}^{\rm up}=A_{\rm p}^{\rm up}$ and
$A_{\rm d}^{\rm down}=A_{\rm p}^{\rm down}$. Thus the measured cross section
ratio does not depend on the incident muon flux or the detector acceptance.

To obtain the ratio of one photon exchange cross sections, 
$\sigma_{\rm d}/\sigma_{\rm p}$, the numbers of events in 
eq.(\ref{eq:ratio}) were replaced by the accumulated weights,
$\sigma/\sigma^{\rm incl}$, to correct for higher order electroweak processes.
These radiative corrections were calculated using the method of
ref. \cite{radcor} as described in ref.\,\cite{nmcf2}. This procedure
includes corrections for the radiative tails of elastic and quasi-elastic
 scattering
as well as for the inelastic radiative tails. For the calculation of the latter
the structure function $F_2$ and the ratio $R$ are needed.
Therefore the extraction of $\sigma_{\rm d}/\sigma_{\rm p}$ was
 performed in an iterative
procedure. The structure function $F_2^{\rm p}$ was fixed to the
parametrisation of NMC, SLAC and BCDMS
data from ref.\,\cite{nmcnewf2} and for $R$ the parametrisation of
 ref.\,\cite{rslac}
was used.
The structure function $F_2^{\rm d}$ was taken as the product of 
$F_2^{\rm p}$
and $\sigma_{\rm d}/\sigma_{\rm p}$, which was
determined from a fit in $x$ and $Q^2$ to
 the
presently measured ratio, assuming $R^{\rm d}=R^{\rm
 p}$.
For the extrapolation of $F_2^{\rm p}$ to $Q^2=0$, the model of Donnachie
 and
Landshoff \cite{dl} was used, while $R$
was assumed to be constant below $Q^2=0.35$ GeV$^2$. 
For the form factor of the nucleon the parametrisation of Gari and Kr\"umpelmann
\cite{gari} was used and for the deuteron the parametrisation of 
~\v{S}varc and Locher \cite{locher}.
The suppression of quasi-elastic scattering in the deuteron
was evaluated
using the results of a calculation of Bernabeu and Pascual \cite{bernabeu}.
The iteration was stopped
when the change in $\sigma_{\rm d}/\sigma_{\rm p}$ was less than 0.1\%
in every $x$ and $Q^2$ bin.

To estimate the error due to radiative corrections the prescription given in
ref.\,\cite{nmcratlong} was followed.
For this estimate the upper and lower bounds of $F_2^{\rm p}$ 
from ref.\,\cite{nmcnewf2}
were used and the uncertainty on $R$ for
$Q^2>0.35$ GeV$^2$ was taken to be that of the parametrisation of
ref.\,\cite{rslac} enlarged by 50\%;
for $Q^2$ below 0.35 GeV$^2$
an error on $R$ of +150\% and --100\% was assumed.
For the form factors and the suppression factor  alternative parametrisations
were used to estimate the associated systematic uncertainties.
To estimate the influence of the functional form chosen to parametrise
$\sigma_{\rm d}/\sigma_{\rm p}$, the procedure was
 repeated with a
parametrisation of $\sigma_{\rm d}/\sigma_{\rm p}$ that was a function
 of $x$ only. The total error due to the radiative corrections was at most
2\% at the smallest $x$.

\subsection{Data selection}

To check the assumption of equal acceptances for the two upstream and the two
 downstream
targets the beam flux ratio
\begin{equation}
\frac{\phi_2}{\phi_1}=\sqrt{\frac{N_{\rm d}^{\rm up}N_{\rm p}^{\rm down}}
{N_{\rm p}^{\rm up}N_{\rm d}^{\rm down}}}
\end{equation}
was used. It
should not depend on any variable characterising the event. Thus, cuts were
applied in order to remove events from kinematic regions where
the flux ratio was not constant, e.g. at the edges of the
distributions of the scattering angle, $\theta$, and the energy transfer, $\nu$.

The time dependence of the detector acceptance
was investigated using the acceptance ratio calculated from
\begin{equation}
\frac{A^{\rm up}}{A^{\rm down}}=\sqrt{\frac{N_{\rm d}^{\rm up}N_{\rm p}^{\rm up}}
{N_{\rm d}^{\rm down}N_{\rm p}^{\rm down}}}
\end{equation}
for consecutive exposures of the two target sets. No significant
time dependence of the ratio was observed except for brief periods where
experimental problems could be identified,
and whose data were discarded.

In this analysis we have included the information from the large drift chambers
(W45)  as described in
ref.\,\cite{nmcratlong} contrary to what was done in the $F_2$ analysis
\cite{nmcf2}.
Thus the measurement of cross section ratios could be extended to larger
scattering angles and hence to larger $Q^2$.
Cuts were applied to the incoming muon distributions such
that the incident flux was identical for the upstream and downstream
targets. Cuts on the interaction vertex position were used to associate the
events to one of the targets.

For events selected by the small $x$ trigger, T14, a cut of
 $x_{\rm min}=0.001$ removed the data around
$x=m_{\rm e}/m_{\rm p}=0.544 \cdot 10^{-3}$,
dominated by elastic
scattering from atomic electrons \cite{seitz,nmcnewli}.
The contamination from these events in the region  with $x>0.001$ was estimated
from a Monte Carlo simulation
to be less than 0.5\% for $x<0.002$ and negligible elsewhere.

The final data samples were obtained from the reconstructed events by applying
the kinematic cuts listed in table \ref{tab:cuts}. The cuts exclude
kinematic regions where higher order electroweak processes dominate,
which are contaminated with muons from hadron decays or have
poor kinematic or vertex
resolution.
In addition, certain ($x,Q^2$) bins at the edge of the acceptance were removed;
in particular, bins were discarded which had only a few events from one
of the targets or if their
area $\Delta x \Delta Q^2$
was reduced by more than
half by the kinematic cuts.

\begin{table}[htb]
{
\begin{center}
\begin{tabular}{|cr||c|c|c|c|c|c||c|} \hline
$E$ &Trigger &$y_{\rm max}$&$p'_{\rm min}$&
$\nu_{\rm min}$&$\nu_{\rm max}$&$\theta_{\rm min}$&$\theta_{\rm max}$&No. of events\\
(GeV) &  &    & (GeV) & (GeV) & (GeV) & (mrad) & (mrad) &after cuts\\
\hline
280 & T1 & 0.9   & 40     & 10  &   --  & 10 & -- & $1.41 \cdot 10^6$ \\
       & T2 & 0.9   & 40     & 15  &   --  &  5 &  17 &$1.75 \cdot 10^6$\\
       &T14 & 0.9   & 40     & 20  &   --  &  3.75 & 14.4 &$0.34 \cdot 10^6$\\
\hline
200 & T1 & 0.9   & 30     & 15  &   --  & 10 & -- &$0.71 \cdot 10^6$\\
       & T2 & 0.9   & 30     & 20  &  160  &  6 &  17 &$0.36 \cdot 10^6$\\
       &T14 & 0.9   & 30     & 20  &   --  &  3.75 &  14.4 &$0.25 \cdot 10^6$\\
\hline
120 & T1 & 0.9   & 20     &  7  &   --  & 14.3 & -- &$1.12 \cdot 10^6$\\
       & T2 & 0.9   & 20     & 10  &   90  &  6 &  17 &$0.31 \cdot 10^6$\\
       &T14 & 0.9   & 20     & 10  &   --  &  3.75 & 14.4 &$0.18 \cdot 10^6$\\
\hline
 90 & T1 & 0.9   & 15     &  5  &   --  & 13 & -- &$1.80 \cdot 10^6$\\
       & T2 & 0.9   & 15     &  5  &   --  &  3 & 17 &$0.18 \cdot 10^6$\\
\hline
\end{tabular}
\end{center}
}
\caption{Kinematic cuts applied to the eleven data sets classified by
energy and trigger. The scattered muon momentum is indicated by $p'$.}
\label{tab:cuts}
\end{table}

The total number of events in the analysis after cuts is $2.7\cdot 10^6$
for hydrogen and $5.7\cdot 10^6$ for deuterium. This is more than double the
number of events in the analysis of ref.\,\cite{nmcratlong}, with the increase
largest at small $x$. About a quarter of the events were used 
in the determination of
$F_2^{\rm d}$, $F_2^{\rm p}$ and $R$ of ref.\,\cite{nmcr}.

\subsection{Further corrections to $\sigma_{\rm d}/\sigma_{\rm p}$  
and systematic errors}

Several corrections were applied to the data in addition to the radiative
corrections discussed earlier.
The finite resolution of the spectrometer caused some vertices 
to be reconstructed outside the targets or even
to be associated 
to the wrong target. To estimate the number of such events, the
longitudinal vertex distributions were fitted in several
intervals of the scattering angle. These fits were used to determine
the optimal vertex cuts and the corrections due to the tails
of the distributions.
These corrections varied between 0.1\% and 1\% and the error was taken as 
half the
value with a minimum of 0.1\%.

The correction due to the finite kinematic resolution of the spectrometer was
taken from a Monte Carlo simulation of the experiment.
This correction was usually much below 1\%
except for $x>0.4$ where it reached several percent for the lowest $Q^2$ bins.
The error was taken to be 30\% of the correction.

In addition, the effects of the HD admixture in the deuterium
and of the mylar in the beam
were taken into account,
yielding changes in the ratio of  about 1\% and 0.3\%, respectively. 
The errors on the ratio due to the uncertainties on these
corrections were less than 0.1\%.

The corrections were taken into account separately for each of the eleven
sets of data listed
in table \ref{tab:cuts}.
For the further analysis the ratios were interpolated to the
centre of each $x$ bin and
the results for each incident energy obtained by taking the
geometrical average \cite{bodek} of the corresponding data sets.

The total systematic error on $\sigma_{\rm d}/\sigma_{\rm p}$ was
 determined by adding
in quadrature the
contributions from the uncertainties on the muon  momenta, on the 
radiative, vertex and smearing corrections
 and on those due to the functional form used to describe the ratio
during the iterations. The latter four were assumed to be fully correlated
between the data sets.
The uncertainties in the hydrogen and deuterium densities and target lengths
led to an additional
 normalisation error which is smaller than 0.1\% (relative normalisation
uncertainties are discussed in section 5).

\section{Results for $R^{\rm d}-R^{\rm p}$}

\begin{figure}[htb]
\begin{center}
\mbox{\epsfig{figure=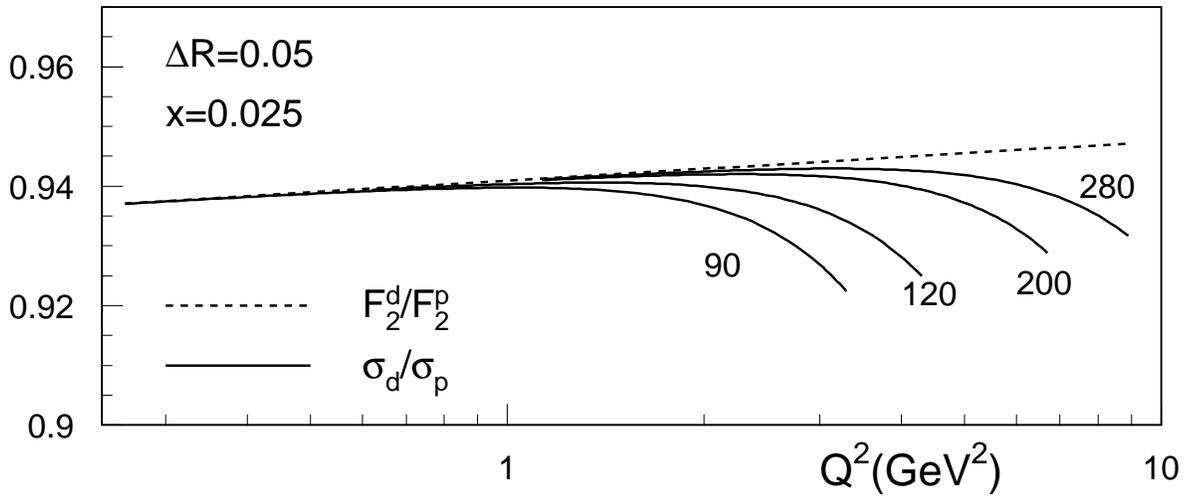,width=\textwidth}
 }
\end{center}
\caption{Comparison of $F_{2}^{\rm d}/F_{2}^{\rm p}$ (dashed line) and
$\sigma_{\rm d}/\sigma_{\rm p}$ (solid lines) for different 
incident muon energies
of 90, 120, 200 and 280~GeV assuming $\Delta R=0.05$.
The ratios are shown as a function of $Q^2$ at $x=0.025$ for the
$Q^2$ range covered by the data at each incident energy.
}
\label{fig:explain}
\end{figure}

The difference  $\Delta R =R^{\rm d}-R^{\rm p}$
was determined following the method described
in ref.\,\cite{nmcdeltar}. From eq.(\ref{eq:sigma}) the cross section
ratio, $\sigma_{\rm d}/\sigma_{\rm p}$, can be related to the structure
function ratio, $F_2^{\rm d}/F_2^{\rm p}$, and
$R^{\rm d}$ and $R^{\rm p}$  through
\begin{equation}
\frac{\sigma_{\rm d}}{\sigma_{\rm p}}(x,Q^2,E)=
\frac{F_2^{\rm d}}
{F_2^{\rm p}}(x,Q^2) \cdot
\frac{1+R^{\rm p}(x,Q^2)}{1+R^{\rm d}(x,Q^2)} \cdot
\frac{1+\varepsilon \cdot R^{\rm d}(x,Q^2)}{1+\varepsilon \cdot R^{\rm
 p}(x,Q^2)}\ .
\label{eq:sigmrat}
\end{equation}
The dependence of the cross section ratio on the incident energy $E$
appears only through the polarisation parameter $\varepsilon$
\begin{equation}
\varepsilon =\left (1+\frac{1}{2}(1-\frac{2m^2}{Q^2})
\cdot \frac{y^2+\frac{Q^2}{E^2}}{1-y-\frac{Q^2}
{4E^2}}\right )^{-1}.
\label{eq:coef}
\end{equation}
This coefficient is always smaller than unity and mainly dependent
on $y=\nu/E$. Expanding eq.(\ref{eq:sigmrat}) one obtains to first order in
$\Delta R$
\begin{equation}
\frac{\sigma_{\rm d}}{\sigma_{\rm p}} \simeq \frac{F_2^{\rm d}}
{F_2^{\rm p}}\left (1-
\frac{1-\varepsilon}{(1+\overline{R})(1+\varepsilon\overline{R})} \Delta R
 \right ),
\label{eq:deltar}
\end{equation}
where $\overline{R}=(R^{\rm d}+R^{\rm p})/2$. Because $\Delta R$ is
 small,
$\sigma_{\rm d}/\sigma_{\rm p}$ is insensitive to $\overline{R}$.
The sensitivity to
$\Delta R$ is largest for those ($x,Q^2$) bins where the range in
$\varepsilon$ is widest.
Fig.\,\ref{fig:explain} shows schematically the behaviour of the cross
section ratio as a function of $Q^2$ for different energies:
the small $Q^2$ (large $\varepsilon$) data at each energy are insensitive
to $\Delta R$, but at large $Q^2$ (small $\varepsilon$) the sensitivity
becomes significant.
Thus the data at large $\varepsilon$ are mainly sensitive to
$F_2^{\rm d}/F_2^{\rm p}$ while $\Delta R$ is determined by the small
$\varepsilon$ data.

We chose to determine a $Q^2$-averaged value of $\Delta R$ for each $x$ bin
separately, fitting a parametrisation of 
$\sigma_{\rm d}/\sigma_{\rm p}$ to the data using
eq.(\ref{eq:sigmrat}). Four parameters were used: $\Delta R$, 
$\overline{R}$ and the two parameters of the function 
$F_2^{\rm d}/F_2^{\rm p}=a_1+a_2 \ln Q^2$.
Previous measurements of $R$ \cite{nmcr,CDHSW,BCDMS} were
included in the fits
to loosely constrain the value of $\overline{R}$.
By fitting eq.(\ref{eq:sigmrat}) to all data in each 
$x$ bin, the $Q^2$-averaged value of $\Delta R$ was determined with higher
accuracy than would have been achieved if the analysis had been restricted
to the regions of overlap in $Q^2$. 
The lowest $x$ bin was
excluded because the range in $\varepsilon$ is too small.
Data at large $x$ were not included because of their low sensitivity to
$\Delta R$. The fits describe the data well
in all $x$ bins with a total $\chi^2$ of 472 for 515 degrees of freedom.

The systematic errors on $\Delta R$ were calculated
by shifting the measured cross section ratios by the
error due to each source separately and repeating the fits. 
All contributions to the error on $\Delta R$ were
then added in quadrature. The dominant contributions were due to the
uncertainties on the radiative
corrections at small $x$, on the normalisation at medium $x$
and on the muon momenta at large $x$.

\begin{figure}[ht]
\begin{center}
\mbox{\epsfig{figure=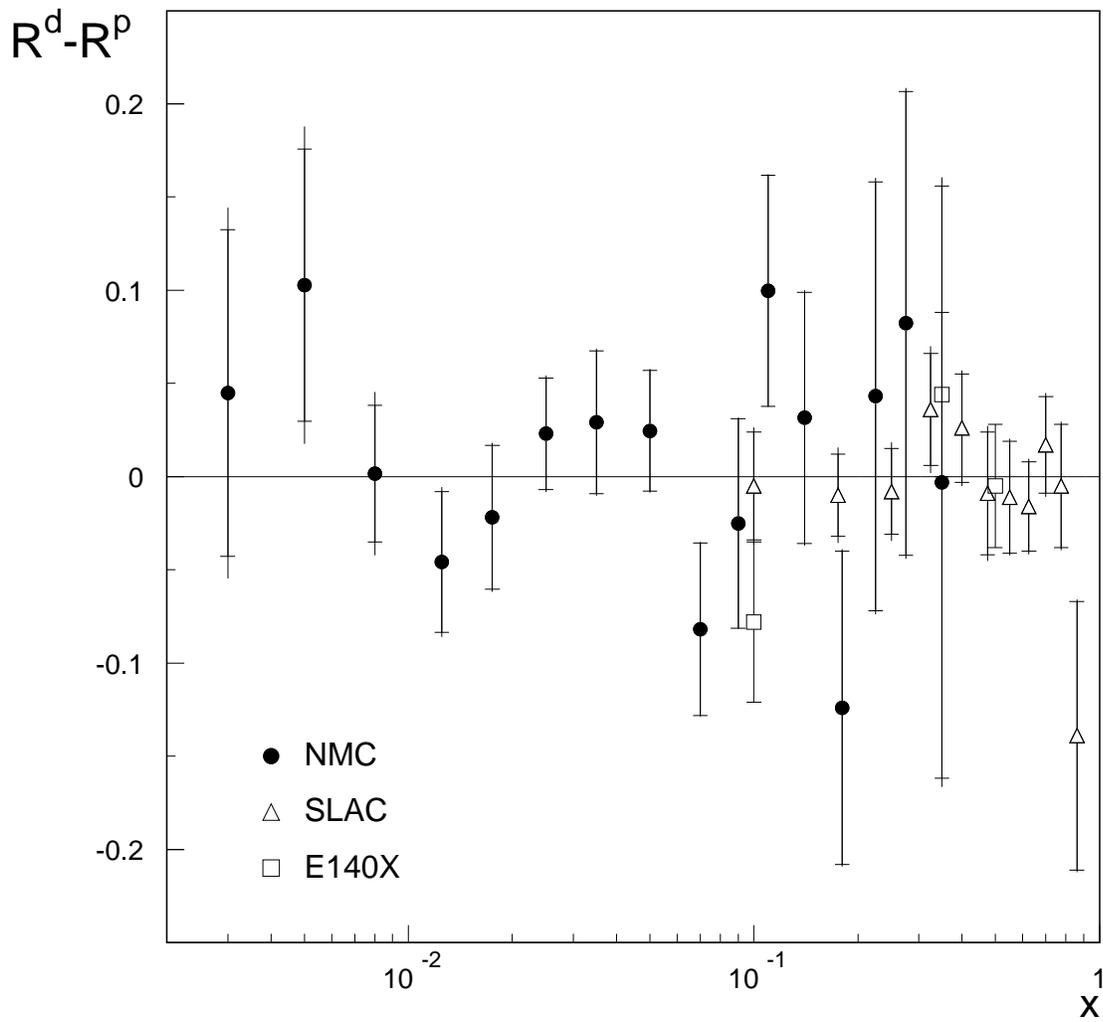,width=\textwidth}}
\end{center}
\caption{
E140X {\protect\cite{e140x}}. The inner
error bars correspond to statistical errors and the full ones to the
quadratic sum of statistical and systematic errors; for the E140X results only 
statistical error bars are given.}
\label{fig:deltar}
\end{figure}

The results for $\Delta R$ are shown in fig.\,\ref{fig:deltar} and listed in
table \ref{tab:deltar}; they cover the range $0.003<x<0.35$.
The $\langle Q^2 \rangle$ given in table \ref{tab:deltar} were evaluated using
weights derived from the sensitivity of each of the data points to $\Delta R$.
The values of $\Delta R$ are small; this is most significant
at small $x$, i.e. small $Q^2$, where $R$ is large
($R\approx 0.3$ for $x\approx 0.01$, $Q^2\approx 2$ GeV$^2$
\cite{nmcr, CDHSW}).  No significant
 $x$ dependence of $\Delta R$ is observed.
Averaging the measurements over $x$ one obtains 
\begin{equation}
\Delta R = 0.004 \, \pm \, 0.012 (\rm{stat})
\, \pm \, 0.011 (\rm{syst.}),
\label{eq:val}
\end{equation}
compatible with zero, at $\langle Q^2 \rangle = $ 5~GeV$^2$.

\begin{table}[htb]
{
\begin{center}
\begin{tabular}{|l|r@{.}l||r@{\ --\ }l|r@{\ --\ }l||r@{}l|c|c|} \hline
\multicolumn{1} {|c|}{$x$} &
\multicolumn{2}{c||} {$\langle Q^2 \rangle$} &
\multicolumn{2}{c|} {$y$ Range} &
\multicolumn{2}{c||} {$\varepsilon$ Range} &
\multicolumn{2} {c|} {$R^{\rm d}-R^{\rm p}$}& Stat. & Syst.\\
    &
\multicolumn{2}{c||} {(GeV$^2$)}  &
\multicolumn{2}{c|} {}     &     
\multicolumn{2}{c||} {}     &     
\multicolumn{2}{c|} {}    & error & error\\
\hline
0.0030&  0&68 &0.31 &0.76 &0.944 &0.461 && 0.045 & 0.088  & 0.047\\
0.0050&  0&99 &0.20 &0.72 &0.989 &0.523 && 0.103 & 0.073  & 0.044\\
0.0080&  1&8 &0.13 &0.80 &0.992 &0.389 &&0.002 & 0.037 & 0.024\\
0.0125&  2&9 &0.08 &0.78 &0.997 &0.420 &--&0.046 & 0.038 & 0.014\\
0.0175&  4&0 &0.09 &0.79 &0.996 &0.389 &--&0.022 & 0.039 & 0.011\\
0.025&   5&0 &0.07 &0.74 &0.998 &0.494 & &0.023 & 0.030 & 0.009\\
0.035&   8&1 &0.06 &0.77 &0.998 &0.444 & &0.029 & 0.038 & 0.009\\
0.050&  11&1 &0.06 &0.74 &0.998 &0.492 & &0.025 & 0.033 & 0.007\\
0.070&  15&3 &0.06 &0.73 &0.998 &0.509 &--&0.082 & 0.046 & 0.007\\
0.090&  20&1 &0.06 &0.75 &0.998 &0.468 &--&0.025 & 0.056 & 0.007\\
0.110&  25&0 &0.06 &0.77 &0.998 &0.431 & &0.100 & 0.062 & 0.008\\
0.140&  27&8 &0.06 &0.73 &0.998 &0.506 &  &0.032 & 0.067 & 0.012\\
0.180&  37&1 &0.06 &0.69 &0.998 &0.558 & --&0.124 & 0.084 & 0.014\\
0.225&  38&8 &0.06 &0.73 &0.998 &0.507 &  &0.043 & 0.115 & 0.022\\
0.275&  57&2 &0.05& 0.64&0.998 &0.642 &  &0.082 & 0.125 & 0.022\\
0.350&  62&7 &0.04& 0.60&0.999 &0.693 & --&0.003 & 0.159 & 0.039\\
\hline
\end{tabular}
\end{center}
}
\caption{Values of $\Delta R=R^{\rm d}-R^{\rm p}$ determined at $x$
 and  $\langle Q^2 \rangle$. The calculation of $\langle Q^2 \rangle$ takes 
into account the sensitivity of the data points to $\Delta R$. In addition the
$y$ and $\varepsilon$ range of the data for each $x$ bin are given.
}\label{tab:deltar}
\end{table}

In fig.\,\ref{fig:deltar} we also show the results on $\Delta R$ at higher $x$
obtained from the SLAC experiment E140X \cite{e140x} and 
from a reanalysis of earlier SLAC data \cite{rslac},
which agree very well with the present values.
The SLAC results were determined
using only kinematic regions where measurements at different energies
overlap in $x$ and $Q^2$.
We have re-evaluated $\Delta R$ from the SLAC data in ref.\,\cite{rslac}
with the method described above and find almost identical results but with 
smaller errors \cite{rome}.
Differences of $R$ have also been measured for several combinations of nuclei
(see refs.\,\cite{nmcdeltar,au,sn,e140x}) and were found to be 
compatible with zero.

\begin{figure}[h]
\begin{center}
\mbox{\epsfig{figure=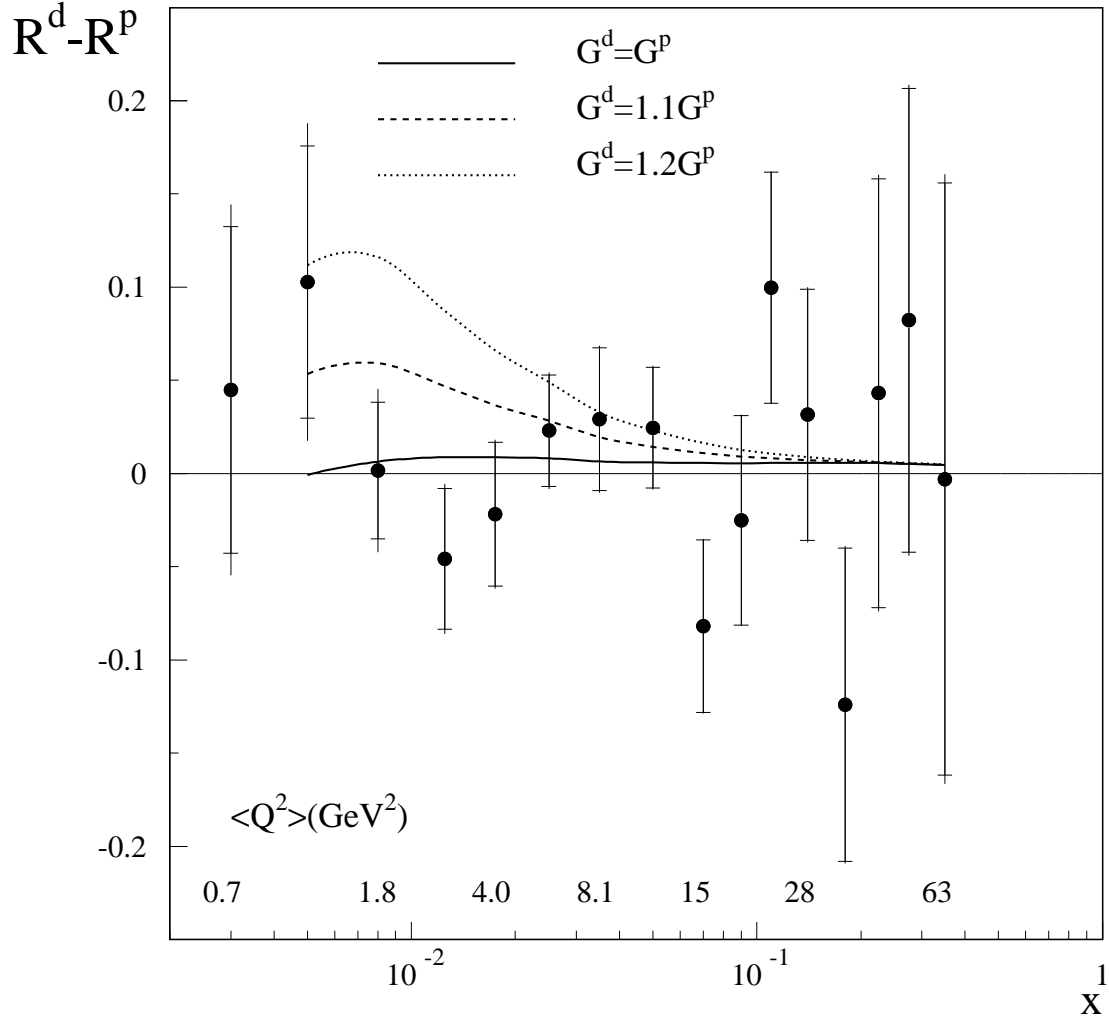,width=\textwidth}}
\end{center}
\caption{Comparison of the present results to predictions from perturbative QCD.
The solid line was calculated using the same gluon distribution for the
proton and the deuteron, the dashed and dotted ones assuming an increase of 
$xG$ for the deuteron of 10\% and 20\%, respectively. 
The numbers given at the bottom of the figure
are the average $Q^2$ values for some $x$ bins.}
\label{fig:rcomp}
\end{figure}

In next to leading order perturbative QCD,
the $x$ and $Q^2$ dependence of $R$ is related \cite{alma}
to that of the singlet structure function
$F_2^{\rm {SI}}$ and the gluon distribution $xG$ through
the longitudinal structure function
\begin{equation}
F_{\rm L}(x,Q^2)=\frac{\alpha_s(Q^2)x^2}{2\pi} \int\limits_x^1 \left(
 \frac{8}{3}
F_2^{\rm {SI}}(w,Q^2) +\frac{40}{9}
wG(w,Q^2)(1-\frac{x}{w}) \right) \frac{\mbox{d}w}{w^3}
\label{eq:altama}
\end{equation}
and
\begin{equation}
R(x,Q^2)=\frac{F_{\rm L}(x,Q^2)+\frac{4M^2x^2}{Q^2} F_2(x,Q^2)}{F_2(x,Q^2)-
F_{\rm L}(x,Q^2)}.
\label{eq:fl}
\end{equation}
For $ x < 0.10 $, the gluon distribution dominates the integral
of eq.(\ref{eq:altama}).  Thus, at small $x$,
$\Delta R$ is sensitive to the difference between the deuteron and proton
gluon distributions. The solid line in fig.\,\ref{fig:rcomp} shows a QCD
prediction for $\Delta R$ assuming equal gluon distributions;
the
values of $F_2$ were taken from ref.\,\cite{nmcnewf2} and the approximation 
was made that $F_2^{\rm {SI}} \approx F_2^{\rm d}$, whereas 
$xG$ was taken from the QCD analysis in ref.\,\cite{qcdH1}.
The dashed and dotted lines in fig.\,\ref{fig:rcomp} were calculated using
a gluon distribution for the deuteron scaled by 1.1 and 1.2, respectively.
Within perturbative QCD this sets a limit on a possible difference
between the proton and the deuteron gluon distributions of about 10\%.
We have not investigated the sensitivity of $\Delta R$ to possible higher
twist effects.

\section{Internal consistency of the data}

We have looked for possible normalisation shifts of the data at 90, 120, 200
and 280 GeV by repeating the calculation of $\Delta R$ with four additional
normalisation parameters in the fit. The $\chi^2$ improvement is not
significant and the normalisation shifts suggested by the fit 
were -0.11\%, -0.01\%,
+0.18\% and +0.06\%, respectively, with typical errors of 0.16\%.
This shows that the internal consistency of the data taken 
over the course of four years is very good.
The compatibility of these shifts with zero
indicates that there is no additional normalisation uncertainty on
$\sigma_{\rm d}/\sigma_{\rm p}$ at the level of 0.15\%. 
The change in $\Delta R$ was everywhere much smaller than the statistical error.

\section{Results for $F_{2}^{\rm d}/F_{2}^{\rm p}$}

Structure function ratios can be determined from the measured cross section
ratios once $\Delta R$ and $\overline{R}$ are known.
As $\Delta R$ is compatible with zero we have taken the structure function
ratio, $F_2^{\rm d}/F_2^{\rm p}$, to be
equal to the cross section ratio,
$\sigma_{\rm d}/\sigma_{\rm p}$.

\begin{figure}[h]
\vspace{1cm}
\begin{sideways}
\begin{minipage}[b]{\textheight}
\begin{center}
\begin{tabular}{c c}
\epsfig{figure=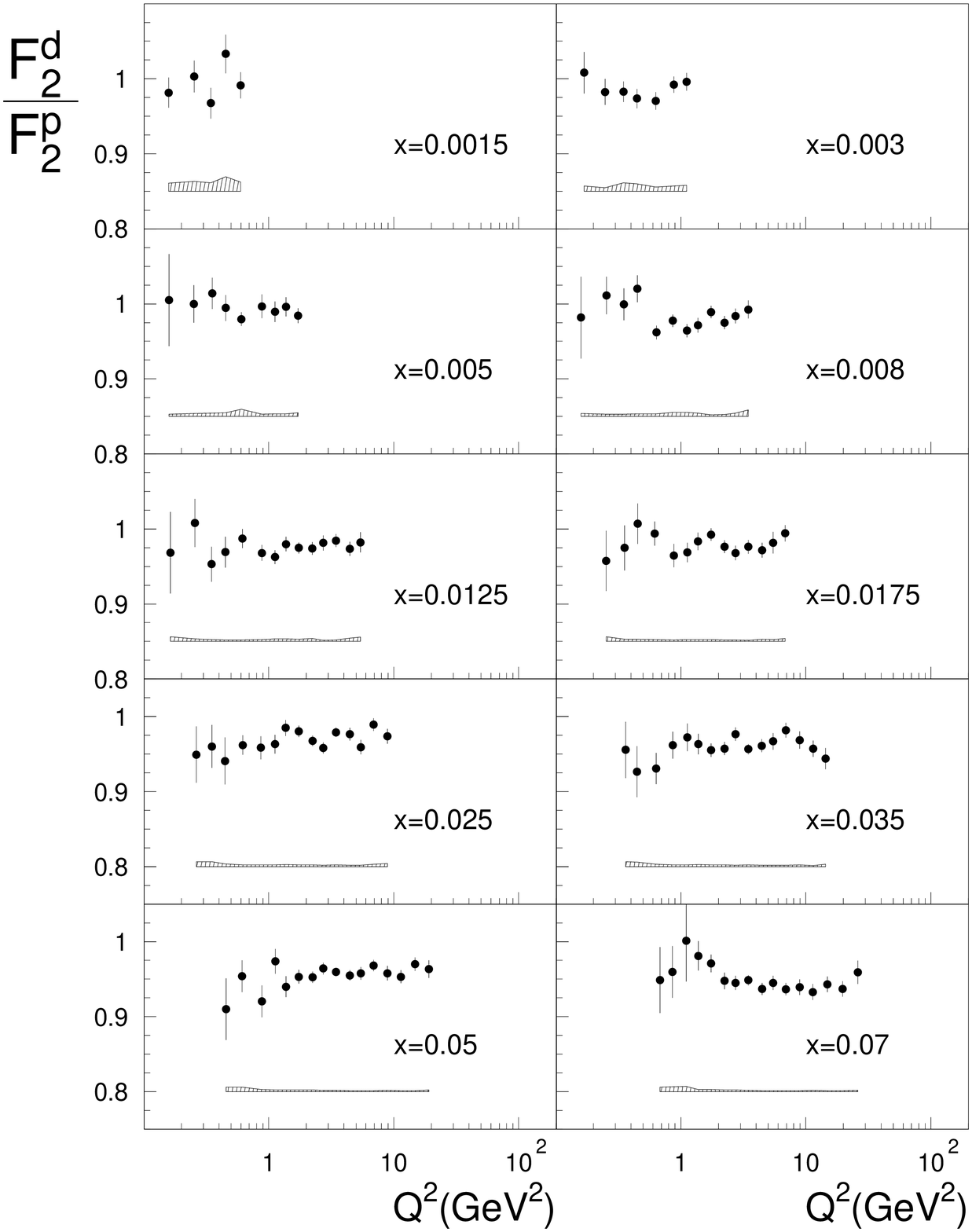,width=.5\textwidth} &
\epsfig{figure=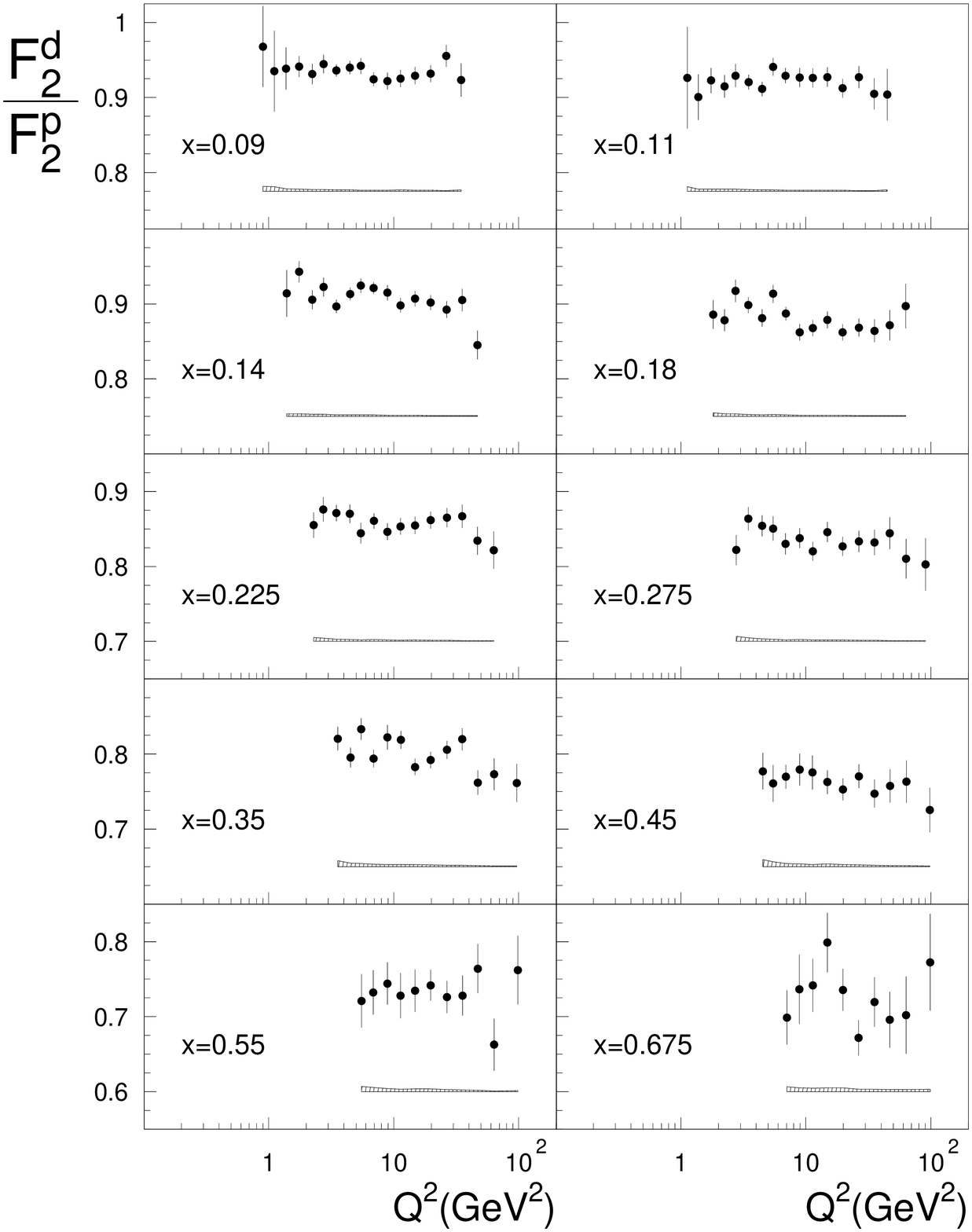,width=.5\textwidth}
\end{tabular}
\end{center}
\caption{The structure function ratio $F_2^{\rm d}/F_2^{\rm p}$
as a function of $Q^2$ for each $x$ bin. The error bars represent the
statistical uncertainties and
the size of the systematic errors is indicated by the bands.}
\label{fig:f2np}
\end{minipage}
\end{sideways}
\end{figure}

\begin{figure}[h]
\begin{center}
\mbox{\epsfig{figure=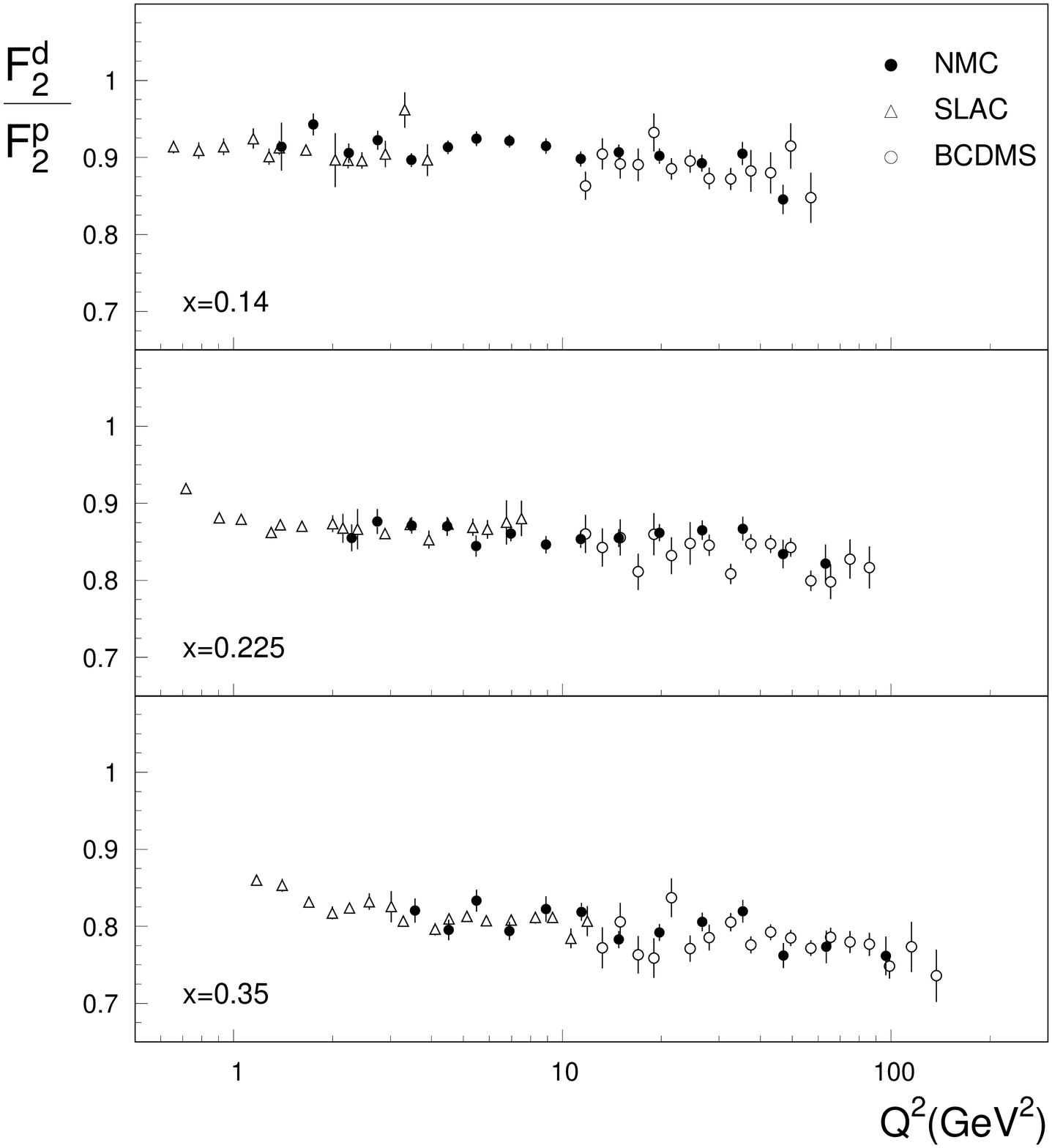,width=\textwidth}}
\end{center}
\caption{Comparison of the present results with the ones from SLAC
{\protect\cite{slacnp}} and BCDMS {\protect\cite{bcdmsnp}}
for selected $x$ bins. The SLAC data were rebinned in $x$ and $Q^2$.
The error bars represent the statistical uncertainties.}
\label{fig:compnp}
\end{figure}

The geometrical average of the data at  the four energies was taken,
and the resulting values for
$F_{2}^{\rm d}/F_{2}^{\rm p}$ in bins of $x$ and $Q^2$
are given in table \ref{tab:f2np}. Fig.\,\ref{fig:f2np} shows the ratio
as a function of $Q^2$ in each $x$ bin. 
The average systematic error is about 0.4\% and almost all
the data have a systematic error smaller than 1\%.
A comparison to results from SLAC \cite{slacnp} and BCDMS \cite{bcdmsnp}
for $F_{2}^{\rm d}/F_{2}^{\rm p}$ is shown in fig.\,\ref{fig:compnp} for
three $x$ bins
and demonstrates good agreement.

\begin{figure}[h]
\begin{center}
\mbox{\epsfig{figure=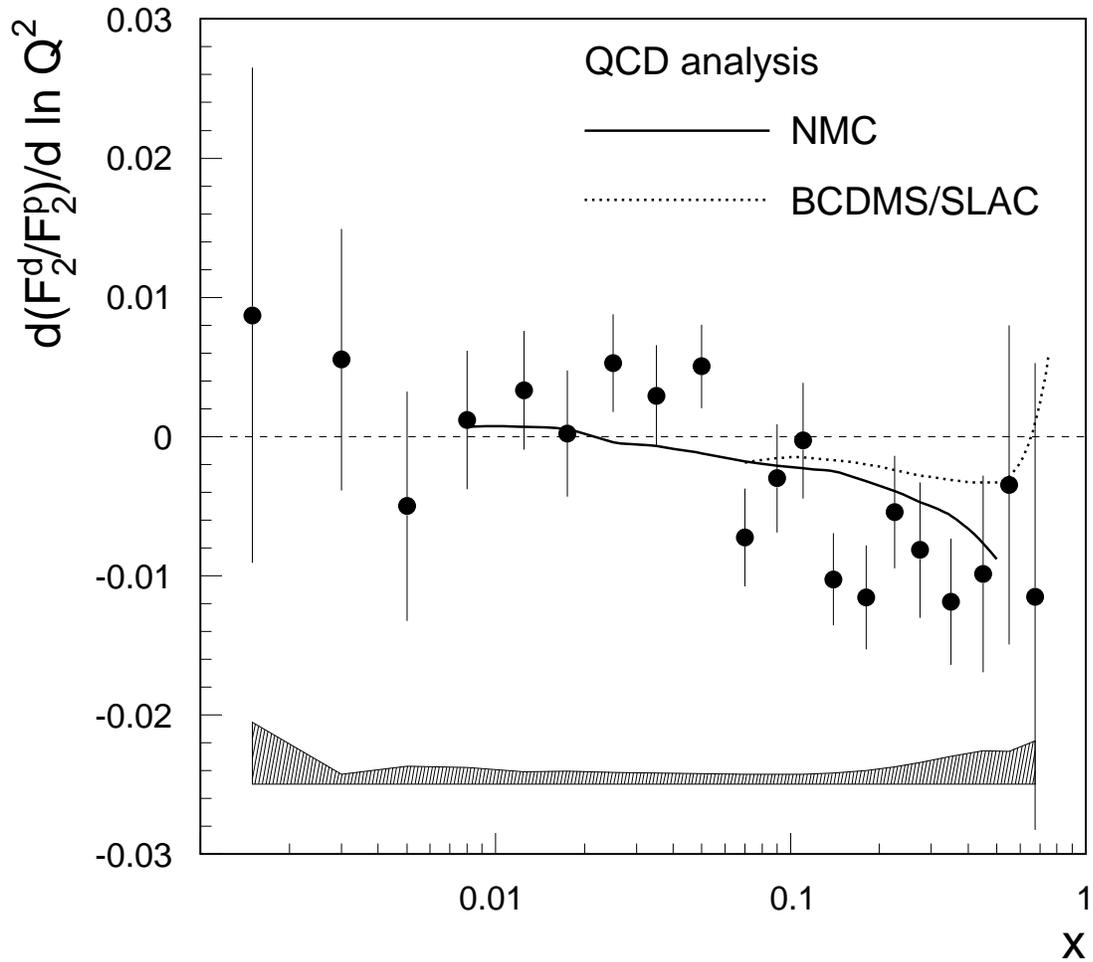,width=\textwidth}}
\end{center}
\caption{The slope parameter
$b_2=\mbox{d}(F_{2}^{\rm d}/F_{2}^{\rm p})/\mbox{d}\ln Q^2$ as a
function of $x$. The error bars represent the statistical uncertainties and
the band indicates the size of the systematic errors.
The curves are results of perturbative QCD calculations, using  NMC structure
function data {\protect\cite{nmcqcd}}
(solid line) and the combined SLAC and BCDMS data {\protect\cite{virch}}
(dashed line). The difference between the two curves is within their systematic
uncertainties.}
\label{fig:slope}
\end{figure}

In most of the $x$ bins the present data cover nearly two decades in
$Q^2$, with little dependence on $Q^2$. 
To investigate possible $Q^2$
dependences, the data were fitted in each $x$ bin 
with a linear function of $\ln Q^2$
\begin{equation}
F_{2}^{\rm d}/F_{2}^{\rm p} =b_1 + b_2 \ln Q^2.
\label{eq:fit}
\end{equation}
Fig.\,\ref{fig:slope}
shows the fitted slope parameter $b_2$ as a function of $x$.
The systematic uncertainties on $b_2$ were calculated by shifting
the ratio by each of its systematic uncertainties and repeating the
fits. The resulting
contributions were added in quadrature. The results of the fits
are given in table \ref{tab:f2npx}.
The total $\chi^2$ of these fits is 202 for 220 degrees of
freedom.

Also shown in fig.\,\ref{fig:slope} are two next to leading order
QCD calculations, including target mass corrections, based on
analyses of the NMC structure function data \cite{nmcqcd} and of the
SLAC and BCDMS data \cite{virch}.
The measured slopes are consistent with these
perturbative QCD calculations although there may be deviations at $x>0.1$
as was suggested in ref.\,\cite{nmcratlong}.
However, in order to investigate the possible interpretation of the
large $x$ data in terms of significant higher twist effects, as was done
in ref.\,\cite{nmcratlong}, a common analysis of the present data and
the SLAC and BCDMS results is required.

The results for $F_{2}^{\rm d}/F_{2}^{\rm p}$ averaged over $Q^2$ are
listed in table \ref{tab:f2npx}.
The statistical and systematic errors are below 0.5\% for most of the
$x$ range. The main contributions to the systematic errors stem from the
uncertainty of the radiative corrections at small $x$ and the uncertainty
in the kinematic resolution and the momentum calibrations at large $x$.
The ratio is consistent with unity at the lowest measured $x$.

\begin{figure}[h]
\begin{center}
\mbox{\epsfig{figure=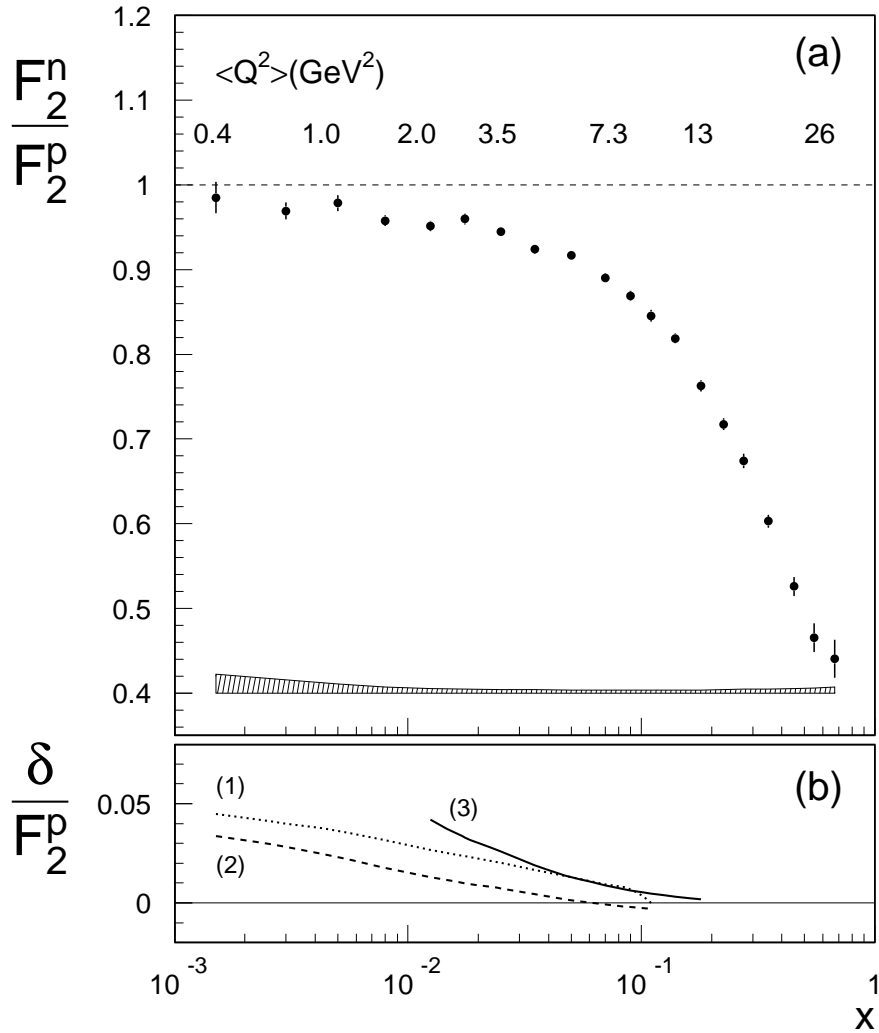,width=.9\textwidth}}
\end{center}
\caption{(a) The structure function ratio $F_{2}^{\rm n}/F_{2}^{\rm p}$
as a function of $x$. The error bars represent the
statistical errors and  the band at the bottom indicates the systematic
uncertainty. The numbers given across the top of the figure are the average
$Q^2$ values.
(b) Model predictions for $\delta/F_2^{\rm p}$, the corrections to 
$F_{2}^{\rm n}/F_{2}^{\rm p}$ 
due to shadowing
in the deuteron, 
from ref.\,{\protect\cite{bk}} line(1), ref.\,{\protect\cite{wmshad}} 
line (2) and ref.\,{\protect\cite{nz}} line (3).}
\label{fig:f2np_x}
\end{figure}

Neglecting nuclear effects in the
deuteron,
 the neutron structure function is
given by $F_2^{\rm n}=2F_2^{\rm d} - F_2^{\rm p}$, and
the ratio of the neutron and proton structure functions by
\begin{equation}
\frac{F_{2}^{\rm n}}{F_{2}^{\rm p}}=
2 \frac{F_{2}^{\rm d}}{F_{2}^{\rm p}} -1
=2\frac{\sigma_{\rm d}}{\sigma_{\rm p}} -1.
\label{eq:f2np}
\end{equation}
The results for $F_{2}^{\rm n}/F_{2}^{\rm p}$
averaged over $Q^2$ calculated according to eq.(\ref{eq:f2np})
are shown in fig.\,\ref{fig:f2np_x}(a). 

The ratio $F_{2}^{\rm n}/F_{2}^{\rm p}$ determined using 
eq.(\ref{eq:f2np}) may deviate significantly from the free nucleon
ratio, $(F_{2}^{\rm n}/F_{2}^{\rm p})_{\rm free}$, due to nuclear 
effects in the deuteron such as those observed in heavier nuclei
(see e.g. ref.\,\cite{emceff}).
At small $x$, $F_2^{\rm d}$ may be reduced
due to shadowing effects, which are also observed in the real photon 
cross section on the deuteron \cite{realg}. The ratio in
fig.\,\ref{fig:f2np_x}(a) shows no clear indication of shadowing, but a
few percent effect is not excluded. 
Near $x=1$ the effect of the extension of the kinematic range to $x=2$ 
in the deuteron
must become apparent, although our data do not extend to large enough
$x$ for this to become visible. Finally, at $x \simeq 0.5$
there may be some depletion of $F_2^{\rm d}$ as observed in 
heavier nuclei (the EMC effect).

Fig.\,\ref{fig:f2np_x}(b) illustrates the size of the shadowing correction
according to various models \cite{bk,wmshad,nz}
expressed as a correction to the ratio, 
$\delta/F_{2}^{\rm p}=(F_{2}^{\rm n}/F_{2}^{\rm p})_{\rm free}
- F_{2}^{\rm n}/F_{2}^{\rm p}$, and 
calculated at the $x$ and $Q^2$ of the present data.
All three models use QCD inspired approaches in the perturbative
(high $Q^2$) region. In addition, the models of refs.\,\cite{bk,wmshad}
introduce a vector meson dominance contribution to shadowing, an important 
dynamical mechanism in the nonperturbative (low $Q^2$) regime. Meson exchange
currents included in the models of refs.\,\cite{wmshad,nz} reduce somewhat 
the size of the shadowing correction.
In the kinematic range covered by the present data the predicted 
shadowing corrections from 
these models are up to 0.02 -- 0.05.

A large number of calculations is available to determine the effects of
Fermi motion and the extension of the kinematic range  
in the deuteron beyond $x=1$ (see e.g. ref.\,\cite{fs}). 
They predict only small
corrections in the $x$ range of the NMC results.
Many models have included effects which have been suggested as the 
source of the EMC effect at $x \simeq 0.5$; for example
refs.\,\cite{wmfermi,bt} suggest that nuclear binding effects could cause 
corrections as large as 0.05 at the largest $x$ of the present data.

\begin{figure}[htb]
\begin{center}
\mbox{\epsfig{figure=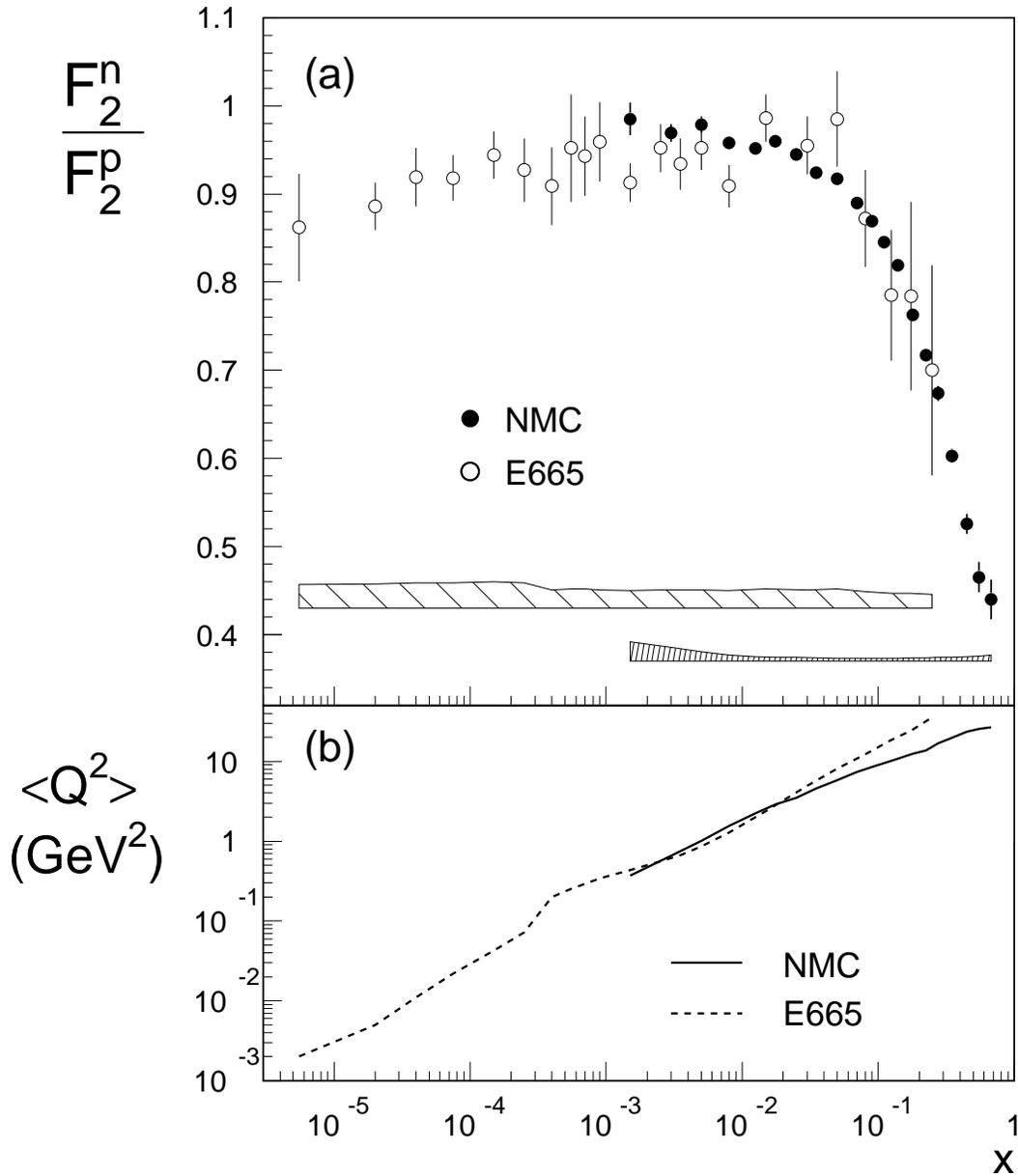,width=\textwidth}}
\end{center}
\caption{(a) Comparison of  the $x$ dependence of the present data for
$F_{2}^{\rm n}/F_{2}^{\rm p}$  with the results from the
E665 collaboration {\protect\cite{e665np}}. The error bars represent the
statistical errors and the bands at the bottom indicate the systematic
uncertainties.\,\,\,\, (b)  The average $Q^2$ of the
E665 and the present data versus $x$.}
\label{fig:f2np_e665}
\end{figure}

In fig.\,\ref{fig:f2np_e665}(a) results for the $x$ dependence of
$F_2^{\rm n}/F_2^{\rm p}$
from the Fermilab E665 collaboration
\cite{e665np} are compared to the present results.
There is fair agreement between the two
experiments while the accuracy of the NMC results is much higher.
As can be seen in fig.\,\ref{fig:f2np_e665}(b) the average $Q^2$ is
quite similar in the region of overlap. 
The E665 results indicate a sizable shadowing effect at very small $x$;
note that these data are at very small $Q^2$.

About half of the present data had been used \cite{gs,gsup}
to determine the Gottfried sum 
\mbox{$S_{\rm G}= \int\limits_0^1 (F_2^{\rm p} -F_2^{\rm
 n})\mbox{d}x/x $} 
where the difference in the integrand was calculated from
\begin{equation}
F_{2}^{\rm p}-F_{2}^{\rm n}=2F_2^{\rm d}\cdot
\frac{1-F_{2}^{\rm d}/F_{2}^{\rm p}}{F_{2}^{\rm d}/F_{2}^{\rm
 p}}\,\,.
\label{eq:pn}
\end{equation}
Using the present values for $F_{2}^{\rm d}/F_{2}^{\rm p}$ and the
parametrisation of $F_2^{\rm d}$ of ref.\,\cite{nmcnewf2}, 
one obtains a contribution to
the Gottfried sum in the interval $0.004<x<0.8$ of
$0.2281 \pm 0.0065$ (stat), at $Q^2=4$ GeV$^2$.
This agrees within statistical errors with
our previously published value \cite{gsup}.

\section{Summary}

The $x$ and $Q^2$ dependence of the cross section ratio,
$\sigma_{\rm d}/\sigma_{\rm p}$, was measured in
deep inelastic muon scattering at
four incident energies with high statistics
and a typical systematic accuracy of 0.5\%.
The results  cover the large kinematic range
$0.001<x<0.8$ and $0.1<Q^2<145$ GeV$^2$ and were obtained from all the
NMC proton and deuteron data.

From the measured cross section ratios, the difference
$R^{\rm d}-R^{\rm p}$ was determined in the
$x$ range from 0.003 to 0.35. It is compatible with zero, 
as expected from perturbative QCD if the proton and deuteron gluon
distributions are equal.

The structure function ratio $F_{2}^{\rm d}/F_{2}^{\rm p}$
shows no $Q^2$ dependence at small $x$ and a small $Q^2$ dependence
compatible with that expected from perturbative QCD at large $x$, 
although higher twist
effects are not excluded.
The ratio $F_{2}^{\rm d}/F_{2}^{\rm p}$ indicates no
sizeable shadowing in the $x$ range covered by our measurements.

\newpage

\begin{table}[h]
\caption{The structure function ratio 
$F_{2}^{\rm d}/F_{2}^{\rm p}$ in bins of $x$ and $Q^2$ with its
statistical and systematic errors.
The systematic error is the quadratic sum of the contributions given 
in  columns 6--10. The contributions are:
the error on the correction due to vertex resolution (VX); the error due
to kinematic resolution (SM); the quadratic sum of errors from radiative
corrections and the functional form of the ratio parametrisation (RC); the error
due to the uncertainty of the incident muon momentum (E); the error due 
to the uncertainty of the scattered muon momentum (E').
The signs of E and E' correspond to an increase in the respective muon energy.}
{\small
\begin{center}
\begin{tabular}{|l|c||c|c|c||rrrrr|} \hline
\multicolumn{1}{|c|}{$x$} &
$Q^2$ & $F_{2}^{\rm d}/F_{2}^{\rm p}$& Stat. &
 Syst.& VX & SM & RC & E & E'\\
 &
(GeV$^2$)    &           & error & error&
 in \%& in \% & in \%& in \%& in \%\\
\hline
0.0015& 0.16&0.9815&0.0203&0.0109& 0.1& 0.0& 1.1& 0.0& 0.0\\
0.0015& 0.25&1.0030&0.0212&0.0134& 0.1& 0.0& 1.3& 0.1& 0.1\\
0.0015& 0.35&0.9675&0.0205&0.0112& 0.2& 0.0& 1.1& 0.0& 0.0\\
0.0015& 0.45&1.0330&0.0258&0.0195& 0.1& 0.0& 1.9& 0.0& 0.0\\
0.0015& 0.60&0.9912&0.0176&0.0121& 0.1& 0.0& 1.2& 0.0& 0.0\\
\hline
0.0030& 0.17&1.0080&0.0277&0.0070& 0.1& 0.0& 0.7& 0.1& 0.1\\
0.0030& 0.25&0.9824&0.0171&0.0047& 0.1& 0.0& 0.5& 0.0& 0.0\\
0.0030& 0.35&0.9825&0.0137&0.0113& 0.2& 0.0& 1.1& 0.0& 0.0\\
0.0030& 0.45&0.9736&0.0129&0.0099& 0.2& 0.0& 1.0& 0.0& 0.0\\
0.0030& 0.63&0.9704&0.0118&0.0057& 0.2& 0.0& 0.6& 0.0& 0.0\\
0.0030& 0.88&0.9921&0.0108&0.0073& 0.1& 0.0& 0.7& 0.0& 0.0\\
0.0030& 1.12&0.9959&0.0116&0.0078& 0.1& 0.0& 0.8& 0.0& 0.0\\
\hline
0.0050& 0.16&1.0050&0.0615&0.0030& 0.2& 0.0& 0.2& 0.0& 0.0\\
0.0050& 0.25&1.0000&0.0250&0.0037& 0.1& 0.0& 0.3&--0.1&--0.1\\
0.0050& 0.35&1.0140&0.0208&0.0043& 0.2& 0.0& 0.4&--0.1&--0.1\\
0.0050& 0.45&0.9945&0.0172&0.0046& 0.2& 0.0& 0.4& 0.0& 0.0\\
0.0050& 0.61&0.9795&0.0092&0.0094& 0.2& 0.0& 0.9& 0.0& 0.0\\
0.0050& 0.88&0.9966&0.0157&0.0032& 0.2& 0.0& 0.2& 0.0& 0.0\\
0.0050& 1.13&0.9893&0.0137&0.0033& 0.2& 0.0& 0.3& 0.0& 0.0\\
0.0050& 1.38&0.9959&0.0128&0.0032& 0.1& 0.0& 0.3& 0.0& 0.0\\
0.0050& 1.71&0.9842&0.0098&0.0048& 0.1& 0.0& 0.5& 0.0& 0.0\\
\hline
0.0080& 0.16&0.9817&0.0547&0.0041& 0.2&--0.3& 0.2& 0.0& 0.0\\
0.0080& 0.25&1.0110&0.0250&0.0030& 0.1& 0.0& 0.2& 0.1& 0.1\\
0.0080& 0.35&0.9993&0.0213&0.0028& 0.2& 0.0& 0.2& 0.0& 0.0\\
0.0080& 0.45&1.0200&0.0180&0.0035& 0.2& 0.0& 0.3&--0.1&--0.1\\
0.0080& 0.64&0.9618&0.0091&0.0036& 0.2& 0.0& 0.3& 0.0& 0.0\\
0.0080& 0.86&0.9775&0.0083&0.0051& 0.2& 0.0& 0.5& 0.0& 0.0\\
0.0080& 1.12&0.9642&0.0088&0.0054& 0.2& 0.0& 0.5& 0.0& 0.0\\
0.0080& 1.37&0.9714&0.0100&0.0045& 0.2& 0.0& 0.4& 0.0& 0.0\\
0.0080& 1.75&0.9891&0.0083&0.0020& 0.1& 0.0& 0.2& 0.0& 0.0\\
0.0080& 2.24&0.9750&0.0086&0.0023& 0.1& 0.0& 0.2& 0.0& 0.0\\
0.0080& 2.73&0.9837&0.0097&0.0042& 0.2& 0.0& 0.4& 0.0& 0.0\\
0.0080& 3.46&0.9924&0.0122&0.0084& 0.3& 0.0& 0.8& 0.0& 0.0\\
\hline
0.0125& 0.16&0.9683&0.0543&0.0065& 0.2&--0.6& 0.2& 0.0& 0.0\\
0.0125& 0.26&1.0080&0.0320&0.0034& 0.2&--0.2& 0.2& 0.0& 0.0\\
0.0125& 0.35&0.9530&0.0233&0.0026& 0.2&--0.1& 0.2& 0.0& 0.0\\
0.0125& 0.45&0.9690&0.0205&0.0022& 0.2& 0.0& 0.2& 0.0& 0.0\\
0.0125& 0.62&0.9872&0.0127&0.0023& 0.2& 0.1& 0.1& 0.0& 0.0\\
0.0125& 0.88&0.9680&0.0108&0.0025& 0.2& 0.0& 0.1& 0.0& 0.0\\
\hline
\end{tabular}
\end{center}
}
\label{tab:f2np}
\end{table}
\begin{table}[h]
\center{Table 3 (continued)}
{\small
\begin{center}
\begin{tabular}{|l|c||c|c|c||rrrrr|} \hline
\multicolumn{1}{|c|}{$x$} &
$Q^2$ & $F_{2}^{\rm d}/F_{2}^{\rm p}$& Stat. &
 Syst.& VX & SM & RC & E & E'\\
 &
(GeV$^2$)    &           & error & error&
 in \%& in \% & in \%& in \%& in \%\\
\hline
0.0125& 1.12&0.9624&0.0092&0.0035& 0.2& 0.0& 0.3& 0.0& 0.0\\
0.0125& 1.37&0.9797&0.0098&0.0035& 0.2& 0.0& 0.3& 0.0& 0.0\\
0.0125& 1.74&0.9747&0.0072&0.0030& 0.2& 0.0& 0.2& 0.0& 0.0\\
0.0125& 2.23&0.9738&0.0085&0.0041& 0.2& 0.0& 0.4& 0.0& 0.0\\
0.0125& 2.74&0.9813&0.0103&0.0016& 0.1& 0.0& 0.1& 0.0& 0.0\\
0.0125& 3.46&0.9844&0.0087&0.0022& 0.1& 0.0& 0.2& 0.0& 0.0\\
0.0125& 4.47&0.9734&0.0095&0.0043& 0.2& 0.0& 0.4& 0.0& 0.0\\
0.0125& 5.41&0.9821&0.0134&0.0058& 0.1& 0.0& 0.6& 0.0& 0.0\\
\hline
0.0175& 0.25&0.9573&0.0402&0.0064& 0.2&--0.6& 0.2& 0.0& 0.0\\
0.0175& 0.35&0.9747&0.0301&0.0033& 0.2&--0.2& 0.2& 0.0& 0.0\\
0.0175& 0.45&1.0070&0.0268&0.0028& 0.2&--0.1& 0.2&--0.1&--0.1\\
0.0175& 0.62&0.9939&0.0161&0.0025& 0.2& 0.1& 0.1& 0.0& 0.0\\
0.0175& 0.88&0.9645&0.0155&0.0021& 0.2& 0.0& 0.1& 0.0& 0.0\\
0.0175& 1.12&0.9685&0.0129&0.0026& 0.2& 0.0& 0.1& 0.0& 0.0\\
0.0175& 1.37&0.9834&0.0119&0.0026& 0.2& 0.0& 0.1& 0.0& 0.0\\
0.0175& 1.75&0.9925&0.0084&0.0025& 0.2& 0.0& 0.2& 0.0& 0.0\\
0.0175& 2.24&0.9763&0.0087&0.0022& 0.2& 0.0& 0.1& 0.0& 0.0\\
0.0175& 2.73&0.9680&0.0098&0.0020& 0.1& 0.0& 0.1& 0.0& 0.0\\
0.0175& 3.48&0.9761&0.0092&0.0018& 0.2& 0.0& 0.1& 0.0& 0.0\\
0.0175& 4.47&0.9716&0.0100&0.0028& 0.1& 0.0& 0.3& 0.0& 0.0\\
0.0175& 5.49&0.9817&0.0143&0.0027& 0.2& 0.0& 0.2& 0.0& 0.0\\
0.0175& 6.83&0.9942&0.0107&0.0040& 0.1& 0.0& 0.4& 0.0& 0.0\\
\hline
0.025& 0.26&0.9493&0.0375&0.0063& 0.2&--0.6& 0.1& 0.0& 0.0\\
0.025& 0.35&0.9601&0.0287&0.0064& 0.2&--0.6& 0.2& 0.0& 0.0\\
0.025& 0.45&0.9408&0.0313&0.0037& 0.2&--0.3& 0.2& 0.0& 0.0\\
0.025& 0.62&0.9620&0.0131&0.0022& 0.2&--0.1& 0.1& 0.0& 0.0\\
0.025& 0.86&0.9585&0.0154&0.0020& 0.2& 0.0& 0.1& 0.0& 0.0\\
0.025& 1.13&0.9631&0.0124&0.0023& 0.2& 0.0& 0.1& 0.0& 0.0\\
0.025& 1.37&0.9849&0.0107&0.0027& 0.3& 0.0& 0.1& 0.0& 0.0\\
0.025& 1.74&0.9802&0.0076&0.0023& 0.2& 0.0& 0.1& 0.0& 0.0\\
0.025& 2.24&0.9677&0.0071&0.0020& 0.2& 0.0& 0.1& 0.0& 0.0\\
0.025& 2.74&0.9581&0.0074&0.0018& 0.2& 0.0& 0.1& 0.0& 0.0\\
0.025& 3.45&0.9790&0.0065&0.0023& 0.1& 0.0& 0.2& 0.0& 0.0\\
0.025& 4.47&0.9764&0.0085&0.0017& 0.2& 0.0& 0.1& 0.0& 0.0\\
0.025& 5.48&0.9592&0.0097&0.0017& 0.2& 0.0& 0.1& 0.0& 0.0\\
0.025& 6.92&0.9893&0.0085&0.0032& 0.2& 0.0& 0.3& 0.0& 0.0\\
0.025& 8.92&0.9738&0.0100&0.0043& 0.1& 0.0& 0.4& 0.0& 0.0\\
\hline
0.035& 0.36&0.9557&0.0375&0.0064& 0.2&--0.6& 0.2&--0.1& 0.1\\
0.035& 0.45&0.9264&0.0340&0.0061& 0.2&--0.6& 0.1& 0.0& 0.0\\
0.035& 0.64&0.9308&0.0207&0.0030& 0.2&--0.2& 0.1& 0.0& 0.0\\
0.035& 0.86&0.9618&0.0179&0.0023& 0.2&--0.1& 0.1& 0.0& 0.0\\
0.035& 1.13&0.9723&0.0183&0.0023& 0.2& 0.0& 0.1& 0.0& 0.0\\
0.035& 1.38&0.9633&0.0138&0.0025& 0.2& 0.0& 0.1& 0.0& 0.0\\
0.035& 1.74&0.9554&0.0089&0.0024& 0.2& 0.0& 0.1& 0.0& 0.0\\
0.035& 2.24&0.9572&0.0087&0.0022& 0.2& 0.0& 0.1& 0.0& 0.0\\
\hline
\end{tabular}
\end{center}
}
\end{table}
\begin{table}[h]
\center{Table 3 (continued)}
{\small
\begin{center}
\begin{tabular}{|l|r@{.}l||c|c|c||rrrrr|} \hline
\multicolumn{1}{|c|}{$x$} &
\multicolumn{2}{c||} {$Q^2$} & $F_{2}^{\rm d}/F_{2}^{\rm p}$& Stat. &
 Syst.& VX & SM & RC & E & E'\\
 &
\multicolumn{2}{c||}{(GeV$^2$)}    &           & error & error&
 in \%& in \% & in \%& in \%& in \%\\
\hline
0.035& 2&74&0.9766&0.0090&0.0019& 0.2& 0.0& 0.1& 0.0& 0.0\\
0.035& 3&46&0.9565&0.0069&0.0021& 0.2& 0.0& 0.1& 0.0& 0.0\\
0.035& 4&45&0.9611&0.0088&0.0016& 0.1& 0.0& 0.1& 0.0& 0.0\\
0.035& 5&47&0.9669&0.0111&0.0018& 0.2& 0.0& 0.1& 0.0& 0.0\\
0.035& 6&92&0.9817&0.0101&0.0018& 0.2& 0.0& 0.1& 0.0& 0.0\\
0.035& 8&96&0.9686&0.0115&0.0021& 0.2& 0.0& 0.1& 0.0& 0.0\\
0.035&11&45&0.9572&0.0107&0.0015& 0.1& 0.0& 0.1& 0.0& 0.0\\
0.035&14&36&0.9439&0.0144&0.0031& 0.1& 0.0& 0.3& 0.0& 0.0\\
\hline
0.050& 0&46&0.9101&0.0412&0.0060& 0.2&--0.6& 0.1&--0.1& 0.1\\
0.050& 0&61&0.9539&0.0214&0.0063& 0.2&--0.6& 0.1&--0.1& 0.1\\
0.050& 0&88&0.9204&0.0213&0.0029& 0.2&--0.2& 0.1& 0.0& 0.0\\
0.050& 1&13&0.9738&0.0167&0.0025& 0.2&--0.1& 0.1& 0.0& 0.0\\
0.050& 1&37&0.9400&0.0138&0.0025& 0.2& 0.0& 0.1& 0.0& 0.0\\
0.050& 1&74&0.9532&0.0092&0.0025& 0.3& 0.0& 0.1& 0.0& 0.0\\
0.050& 2&25&0.9526&0.0077&0.0023& 0.2& 0.0& 0.1& 0.0& 0.0\\
0.050& 2&74&0.9642&0.0077&0.0020& 0.2& 0.0& 0.1& 0.0& 0.0\\
0.050& 3&46&0.9597&0.0058&0.0017& 0.2& 0.0& 0.1& 0.0& 0.0\\
0.050& 4&46&0.9551&0.0069&0.0015& 0.1& 0.0& 0.1& 0.0& 0.0\\
0.050& 5&46&0.9577&0.0085&0.0015& 0.1& 0.0& 0.1& 0.0& 0.0\\
0.050& 6&90&0.9682&0.0076&0.0015& 0.1& 0.0& 0.1& 0.0& 0.0\\
0.050& 8&93&0.9578&0.0097&0.0018& 0.2& 0.0& 0.1& 0.0& 0.0\\
0.050&11&44&0.9532&0.0087&0.0014& 0.1& 0.0& 0.0& 0.0& 0.0\\
0.050&14&82&0.9698&0.0090&0.0015& 0.1& 0.0& 0.1& 0.0& 0.0\\
0.050&19&19&0.9635&0.0119&0.0022& 0.1& 0.0& 0.2& 0.0& 0.0\\
\hline
0.070& 0&68&0.9488&0.0438&0.0063& 0.2&--0.6& 0.1&--0.1& 0.1\\
0.070& 0&86&0.9595&0.0344&0.0063& 0.2&--0.6& 0.1&--0.1& 0.1\\
0.070& 1&11&1.0010&0.0542&0.0068& 0.2&--0.6& 0.1&--0.1&--0.1\\
0.070& 1&38&0.9810&0.0197&0.0028& 0.2&--0.1& 0.1& 0.0& 0.1\\
0.070& 1&74&0.9710&0.0117&0.0027& 0.3& 0.0& 0.1& 0.0& 0.0\\
0.070& 2&24&0.9477&0.0111&0.0023& 0.2& 0.0& 0.1& 0.0& 0.0\\
0.070& 2&75&0.9449&0.0096&0.0023& 0.2&--0.1& 0.1& 0.0& 0.0\\
0.070& 3&47&0.9486&0.0069&0.0019& 0.2& 0.0& 0.1& 0.0& 0.0\\
0.070& 4&47&0.9370&0.0079&0.0016& 0.2& 0.0& 0.0& 0.0& 0.0\\
0.070& 5&47&0.9450&0.0095&0.0014& 0.1& 0.0& 0.0& 0.0& 0.0\\
0.070& 6&91&0.9367&0.0082&0.0015& 0.2& 0.0& 0.0& 0.0& 0.0\\
0.070& 8&91&0.9394&0.0105&0.0015& 0.2& 0.0& 0.0& 0.0& 0.0\\
0.070&11&40&0.9328&0.0105&0.0018& 0.2& 0.0& 0.0& 0.0& 0.0\\
0.070&14&89&0.9432&0.0103&0.0014& 0.1& 0.0& 0.1& 0.0& 0.0\\
0.070&19&63&0.9371&0.0103&0.0013& 0.1& 0.0& 0.1& 0.0& 0.0\\
0.070&26&07&0.9592&0.0157&0.0018& 0.1& 0.0& 0.2& 0.0& 0.0\\
\hline
0.090& 0&90&0.9678&0.0540&0.0065& 0.2&--0.6& 0.1&--0.1& 0.1\\
0.090& 1&11&0.9351&0.0537&0.0062& 0.2&--0.6& 0.1&--0.1& 0.1\\
0.090& 1&38&0.9385&0.0281&0.0032& 0.3&--0.1& 0.1&--0.1& 0.1\\
0.090& 1&76&0.9413&0.0140&0.0028& 0.3& 0.0& 0.1&--0.1& 0.1\\
0.090& 2&24&0.9313&0.0136&0.0025& 0.2& 0.0& 0.1& 0.0& 0.1\\
0.090& 2&75&0.9445&0.0124&0.0024& 0.2&--0.1& 0.1& 0.0& 0.0\\
0.090& 3&49&0.9360&0.0082&0.0021& 0.2&--0.1& 0.1& 0.0& 0.0\\
0.090& 4&47&0.9397&0.0089&0.0018& 0.2&--0.1& 0.0& 0.0& 0.0\\
\hline
\end{tabular}
\end{center}
}
\end{table}
\begin{table}[h]
\center{Table 3 (continued)}
{\small
\begin{center}
\begin{tabular}{|l|r@{.}l||c|c|c||rrrrr|} \hline
\multicolumn{1}{|c|}{$x$} &
\multicolumn{2}{c||} {$Q^2$} & $F_{2}^{\rm d}/F_{2}^{\rm p}$& Stat. &
 Syst.& VX & SM & RC & E & E'\\
 &
\multicolumn{2}{c||}{(GeV$^2$)}    &           & error & error&
 in \%& in \% & in \%& in \%& in \%\\
\hline
0.090& 5&46&0.9420&0.0106&0.0016& 0.1&--0.1& 0.0& 0.0& 0.0\\
0.090& 6&91&0.9245&0.0092&0.0016& 0.2& 0.0& 0.0& 0.0& 0.0\\
0.090& 8&92&0.9218&0.0114&0.0014& 0.1& 0.0& 0.0& 0.0& 0.0\\
0.090&11&37&0.9254&0.0115&0.0017& 0.2& 0.0& 0.0& 0.0& 0.0\\
0.090&14&87&0.9291&0.0116&0.0014& 0.1& 0.0& 0.0& 0.0& 0.0\\
0.090&19&74&0.9319&0.0114&0.0013& 0.1& 0.0& 0.1& 0.0& 0.0\\
0.090&26&36&0.9554&0.0147&0.0012& 0.1& 0.0& 0.1& 0.0& 0.0\\
0.090&34&74&0.9233&0.0222&0.0017& 0.1& 0.0& 0.2& 0.0& 0.0\\
\hline
0.110& 1&13&0.9264&0.0680&0.0063& 0.2&--0.6& 0.1&--0.1& 0.2\\
0.110& 1&38&0.9005&0.0306&0.0031& 0.3&--0.1& 0.1&--0.1& 0.1\\
0.110& 1&75&0.9227&0.0169&0.0031& 0.3& 0.0& 0.1&--0.1& 0.1\\
0.110& 2&24&0.9150&0.0151&0.0026& 0.3& 0.0& 0.1&--0.1& 0.1\\
0.110& 2&74&0.9292&0.0153&0.0027& 0.3& 0.0& 0.1& 0.0& 0.1\\
0.110& 3&49&0.9205&0.0097&0.0022& 0.2&--0.1& 0.1& 0.0& 0.1\\
0.110& 4&47&0.9114&0.0098&0.0020& 0.2&--0.1& 0.0&--0.1& 0.1\\
0.110& 5&46&0.9409&0.0117&0.0018& 0.2&--0.1& 0.0& 0.0& 0.1\\
0.110& 6&90&0.9291&0.0102&0.0017& 0.2& 0.0& 0.0& 0.0& 0.0\\
0.110& 8&92&0.9266&0.0127&0.0014& 0.1& 0.0& 0.0& 0.0& 0.0\\
0.110&11&37&0.9263&0.0126&0.0017& 0.2& 0.0& 0.0& 0.0& 0.0\\
0.110&14&85&0.9272&0.0130&0.0015& 0.2& 0.0& 0.0& 0.0& 0.0\\
0.110&19&74&0.9123&0.0124&0.0012& 0.1& 0.0& 0.0& 0.0& 0.0\\
0.110&26&52&0.9272&0.0147&0.0011& 0.1& 0.0& 0.0& 0.0& 0.0\\
0.110&35&32&0.9050&0.0209&0.0011& 0.1& 0.0& 0.1& 0.0& 0.0\\
0.110&44&94&0.9039&0.0345&0.0019& 0.1& 0.0& 0.2& 0.0& 0.0\\
\hline
0.140& 1&40&0.9140&0.0310&0.0037& 0.3& 0.1& 0.1&--0.2& 0.2\\
0.140& 1&75&0.9427&0.0143&0.0036& 0.3& 0.0& 0.1&--0.2& 0.2\\
0.140& 2&24&0.9056&0.0127&0.0030& 0.3& 0.0& 0.1&--0.1& 0.1\\
0.140& 2&74&0.9223&0.0125&0.0028& 0.3& 0.0& 0.1&--0.1& 0.1\\
0.140& 3&47&0.8966&0.0090&0.0023& 0.2& 0.0& 0.1&--0.1& 0.1\\
0.140& 4&48&0.9132&0.0087&0.0022& 0.2&--0.1& 0.0&--0.1& 0.1\\
0.140& 5&47&0.9242&0.0095&0.0020& 0.2&--0.1& 0.0&--0.1& 0.1\\
0.140& 6&90&0.9212&0.0081&0.0019& 0.2&--0.1& 0.0&--0.1& 0.1\\
0.140& 8&92&0.9147&0.0100&0.0016& 0.1& 0.0& 0.0& 0.0& 0.1\\
0.140&11&37&0.8981&0.0097&0.0016& 0.2& 0.0& 0.0& 0.0& 0.0\\
0.140&14&84&0.9068&0.0101&0.0017& 0.2& 0.0& 0.0& 0.0& 0.0\\
0.140&19&76&0.9018&0.0098&0.0013& 0.1& 0.0& 0.0& 0.0& 0.0\\
0.140&26&55&0.8924&0.0110&0.0011& 0.1& 0.0& 0.0& 0.0& 0.0\\
0.140&35&28&0.9050&0.0149&0.0011& 0.1& 0.0& 0.0& 0.0& 0.0\\
0.140&46&95&0.8453&0.0191&0.0010& 0.1& 0.0& 0.1& 0.0& 0.0\\
\hline
0.180& 1&82&0.8859&0.0193&0.0047& 0.3& 0.3& 0.1&--0.2& 0.3\\
0.180& 2&24&0.8781&0.0146&0.0037& 0.3& 0.1& 0.0&--0.2& 0.2\\
0.180& 2&75&0.9173&0.0150&0.0033& 0.3& 0.0& 0.0&--0.2& 0.2\\
0.180& 3&47&0.8983&0.0106&0.0027& 0.2&--0.1& 0.0&--0.1& 0.1\\
0.180& 4&47&0.8811&0.0117&0.0022& 0.2&--0.1& 0.0&--0.1& 0.1\\
0.180& 5&49&0.9134&0.0120&0.0025& 0.2&--0.1& 0.0&--0.1& 0.1\\
0.180& 6&92&0.8869&0.0093&0.0022& 0.2&--0.1& 0.0&--0.1& 0.1\\
\hline
\end{tabular}
\end{center}
}
\end{table}
\begin{table}[h]
\center{Table 3 (continued)}
{\small
\begin{center}
\begin{tabular}{|l|r@{.}l||c|c|c||rrrrr|} \hline
\multicolumn{1}{|c|}{$x$} &
\multicolumn{2}{c||} {$Q^2$} & $F_{2}^{\rm d}/F_{2}^{\rm p}$& Stat. &
 Syst.& VX & SM & RC & E & E'\\
 &
\multicolumn{2}{c||}{(GeV$^2$)}    &           & error & error&
 in \%& in \% & in \%& in \%& in \%\\
\hline
0.180& 8&93&0.8622&0.0108&0.0018& 0.2&--0.1& 0.0&--0.1& 0.1\\
0.180&11&37&0.8676&0.0109&0.0017& 0.2& 0.0& 0.0&--0.1& 0.1\\
0.180&14&85&0.8787&0.0114&0.0018& 0.2& 0.0& 0.0&--0.1& 0.1\\
0.180&19&75&0.8620&0.0108&0.0014& 0.1& 0.0& 0.0&--0.1& 0.1\\
0.180&26&61&0.8684&0.0122&0.0013& 0.1& 0.0& 0.0& 0.0& 0.1\\
0.180&35&37&0.8641&0.0153&0.0011& 0.1& 0.0& 0.0& 0.0& 0.1\\
0.180&47&01&0.8715&0.0202&0.0010& 0.1& 0.0& 0.0& 0.0& 0.0\\
0.180&63&04&0.8970&0.0297&0.0011& 0.1& 0.0& 0.1& 0.0& 0.0\\
\hline
0.225& 2&28&0.8552&0.0172&0.0055& 0.3& 0.3& 0.0&--0.3& 0.3\\
0.225& 2&74&0.8761&0.0161&0.0044& 0.2& 0.1& 0.0&--0.3& 0.3\\
0.225& 3&48&0.8714&0.0109&0.0032& 0.2& 0.0& 0.0&--0.2& 0.2\\
0.225& 4&47&0.8702&0.0121&0.0027& 0.2&--0.1& 0.0&--0.1& 0.2\\
0.225& 5&47&0.8445&0.0137&0.0023& 0.2&--0.1& 0.0&--0.1& 0.1\\
0.225& 6&97&0.8607&0.0101&0.0024& 0.2&--0.1& 0.0&--0.1& 0.2\\
0.225& 8&93&0.8464&0.0112&0.0021& 0.2&--0.1& 0.0&--0.1& 0.1\\
0.225&11&38&0.8534&0.0112&0.0018& 0.2& 0.0& 0.0&--0.1& 0.1\\
0.225&14&85&0.8549&0.0116&0.0019& 0.2& 0.0& 0.0&--0.1& 0.1\\
0.225&19&74&0.8617&0.0112&0.0017& 0.1& 0.0& 0.0&--0.1& 0.1\\
0.225&26&64&0.8651&0.0126&0.0015& 0.1& 0.0& 0.0&--0.1& 0.1\\
0.225&35&42&0.8670&0.0155&0.0013& 0.1& 0.0& 0.0&--0.1& 0.1\\
0.225&46&95&0.8343&0.0186&0.0011& 0.1& 0.0& 0.0& 0.0& 0.1\\
0.225&63&23&0.8218&0.0248&0.0010& 0.1& 0.0& 0.0& 0.0& 0.1\\
\hline
0.275& 2&78&0.8219&0.0203&0.0068& 0.2& 0.5& 0.0&--0.4& 0.5\\
0.275& 3&46&0.8637&0.0155&0.0051& 0.2& 0.1& 0.0&--0.4& 0.4\\
0.275& 4&47&0.8542&0.0141&0.0033& 0.2& 0.0& 0.0&--0.2& 0.2\\
0.275& 5&47&0.8505&0.0163&0.0030& 0.2&--0.1& 0.0&--0.2& 0.2\\
0.275& 6&89&0.8302&0.0142&0.0023& 0.2&--0.1& 0.0&--0.1& 0.2\\
0.275& 8&94&0.8377&0.0134&0.0025& 0.2& 0.0& 0.0&--0.2& 0.2\\
0.275&11&37&0.8202&0.0128&0.0020& 0.2& 0.0& 0.0&--0.1& 0.1\\
0.275&14&87&0.8459&0.0135&0.0022& 0.2& 0.0& 0.0&--0.1& 0.1\\
0.275&19&74&0.8269&0.0127&0.0019& 0.1& 0.0& 0.0&--0.1& 0.1\\
0.275&26&60&0.8334&0.0140&0.0017& 0.1& 0.0& 0.0&--0.1& 0.1\\
0.275&35&43&0.8320&0.0169&0.0014& 0.1& 0.0& 0.0&--0.1& 0.1\\
0.275&46&98&0.8444&0.0214&0.0013& 0.1& 0.0& 0.0&--0.1& 0.1\\
0.275&63&48&0.8104&0.0265&0.0010& 0.1& 0.0& 0.0& 0.0& 0.1\\
0.275&90&68&0.8027&0.0349&0.0010& 0.1& 0.0& 0.1& 0.0& 0.0\\
\hline
0.350& 3&57&0.8205&0.0158&0.0080& 0.2& 0.5& 0.0&--0.6& 0.6\\
0.350& 4&51&0.7953&0.0134&0.0045& 0.2& 0.2& 0.0&--0.3& 0.4\\
0.350& 5&48&0.8332&0.0143&0.0041& 0.2& 0.0& 0.0&--0.3& 0.3\\
0.350& 6&90&0.7939&0.0120&0.0031& 0.2& 0.0& 0.0&--0.2& 0.3\\
0.350& 8&91&0.8222&0.0165&0.0026& 0.1& 0.0& 0.0&--0.2& 0.2\\
\hline
\end{tabular}
\end{center}
}
\end{table}
\begin{table}[h]
\center{Table 3 (continued)}
{\small
\begin{center}
\begin{tabular}{|l|r@{.}l||c|c|c||rrrrr|} \hline
\multicolumn{1}{|c|}{$x$} &
\multicolumn{2}{c||} {$Q^2$} & $F_{2}^{\rm d}/F_{2}^{\rm p}$& Stat. &
 Syst.& VX & SM & RC & E & E'\\
 &
\multicolumn{2}{c||}{(GeV$^2$)}    &           & error & error&
 in \%& in \% & in \%& in \%& in \%\\
\hline
0.350&11&43&0.8188&0.0118&0.0028& 0.2& 0.0& 0.0&--0.2& 0.2\\
0.350&14&85&0.7828&0.0113&0.0025& 0.2& 0.0& 0.0&--0.2& 0.2\\
0.350&19&74&0.7920&0.0109&0.0023& 0.1& 0.0& 0.0&--0.2& 0.2\\
0.350&26&63&0.8055&0.0120&0.0020& 0.1& 0.0& 0.0&--0.1& 0.2\\
0.350&35&45&0.8197&0.0147&0.0018& 0.1& 0.0& 0.0&--0.1& 0.1\\
0.350&47&12&0.7620&0.0162&0.0013& 0.1& 0.0& 0.0&--0.1& 0.1\\
0.350&63&52&0.7732&0.0213&0.0011& 0.1& 0.0& 0.0& 0.0& 0.1\\
0.350&96&35&0.7614&0.0253&0.0010& 0.1& 0.0& 0.1& 0.0& 0.1\\
\hline
0.450& 4&54&0.7771&0.0242&0.0092& 0.2& 0.7& 0.0&--0.7& 0.7\\
0.450& 5&47&0.7608&0.0245&0.0066& 0.2& 0.4& 0.0&--0.5& 0.6\\
0.450& 6&93&0.7698&0.0161&0.0044& 0.2& 0.2& 0.0&--0.3& 0.4\\
0.450& 8&92&0.7793&0.0211&0.0035& 0.1& 0.1& 0.0&--0.3& 0.3\\
0.450&11&34&0.7754&0.0227&0.0026& 0.1& 0.1& 0.0&--0.2& 0.2\\
0.450&14&89&0.7626&0.0158&0.0036& 0.2& 0.2& 0.0&--0.3& 0.3\\
0.450&19&77&0.7529&0.0146&0.0027& 0.1& 0.1& 0.0&--0.2& 0.2\\
0.450&26&64&0.7705&0.0160&0.0025& 0.1& 0.1& 0.0&--0.2& 0.2\\
0.450&35&50&0.7474&0.0187&0.0020& 0.1& 0.1& 0.0&--0.1& 0.2\\
0.450&47&26&0.7575&0.0222&0.0015& 0.1& 0.0& 0.0&--0.1& 0.1\\
0.450&63&65&0.7632&0.0281&0.0012& 0.1& 0.0& 0.0&--0.1& 0.1\\
0.450&98&05&0.7254&0.0296&0.0010& 0.1& 0.0& 0.1& 0.0& 0.1\\
\hline
0.550& 5&53&0.7209&0.0356&0.0072& 0.2& 0.8& 0.1&--0.4& 0.4\\
0.550& 6&88&0.7323&0.0296&0.0055& 0.2& 0.5& 0.1&--0.3& 0.4\\
0.550& 8&91&0.7442&0.0280&0.0042& 0.1& 0.4& 0.1&--0.2& 0.3\\
0.550&11&34&0.7280&0.0300&0.0032& 0.1& 0.3& 0.1&--0.2& 0.2\\
0.550&14&85&0.7345&0.0282&0.0037& 0.2& 0.4& 0.1&--0.2& 0.2\\
0.550&19&74&0.7419&0.0205&0.0037& 0.1& 0.3& 0.1&--0.2& 0.2\\
0.550&26&64&0.7263&0.0216&0.0029& 0.1& 0.2& 0.0&--0.2& 0.2\\
0.550&35&57&0.7281&0.0267&0.0023& 0.1& 0.1& 0.0&--0.2& 0.2\\
0.550&47&16&0.7641&0.0331&0.0021& 0.1& 0.1& 0.0&--0.1& 0.2\\
0.550&63&56&0.6626&0.0345&0.0010& 0.1& 0.0& 0.0&--0.1& 0.1\\
0.550&98&82&0.7622&0.0458&0.0012& 0.1& 0.0& 0.0&--0.1& 0.1\\
\hline
0.675& 7&04&0.6989&0.0361&0.0067& 0.1& 0.7& 0.2& 0.4&--0.4\\
0.675& 8&88&0.7365&0.0465&0.0053& 0.1& 0.6& 0.2& 0.3&--0.3\\
0.675&11&36&0.7418&0.0353&0.0046& 0.1& 0.6& 0.2& 0.1&--0.1\\
0.675&14&85&0.7988&0.0395&0.0051& 0.2& 0.6& 0.1& 0.0&--0.1\\
0.675&19&79&0.7357&0.0281&0.0049& 0.2& 0.6& 0.1&--0.1& 0.1\\
0.675&26&49&0.6717&0.0235&0.0034& 0.1& 0.4& 0.0&--0.2& 0.2\\
0.675&35&40&0.7194&0.0330&0.0033& 0.1& 0.3& 0.0&--0.2& 0.3\\
0.675&47&03&0.6959&0.0373&0.0026& 0.1& 0.1& 0.1&--0.2& 0.3\\
0.675&63&53&0.7020&0.0513&0.0029& 0.1& 0.0& 0.1&--0.2& 0.3\\
0.675&99&03&0.7724&0.0645&0.0034& 0.1& 0.0& 0.2&--0.2& 0.3\\
\hline
\end{tabular}
\end{center}
}
\end{table}

\begin{table}[htb]
{
\begin{center}
\begin{tabular}{|l|r@{.}l|c|r|} \hline
\multicolumn{1}{|c|}{$x$} &
\multicolumn{2}{c|} {$\langle Q^2 \rangle$} &
$F_{2}^{\rm d}/F_{2}^{\rm p} \pm \Delta^{\rm
 stat}
\pm \Delta^{\rm syst}$&
\multicolumn{1}{c|}{$b_2\pm \Delta^{\rm stat}
\pm \Delta^{\rm syst}$}\\
 &
\multicolumn{2}{c|} {(GeV$^2$)} &&\\\hline
0.0015 & 0&37 & 0.9925   $\pm$   0.0092   $\pm$   0.0110
& 0.0087   $\pm$   0.0178    $\pm$   0.0045 \\
0.0030 & 0&66 & 0.9846   $\pm$   0.0051   $\pm$   0.0078
& 0.0055   $\pm$   0.0094    $\pm$   0.0007 \\
0.0050 & 1&0 & 0.9894   $\pm$   0.0047   $\pm$   0.0053
& --0.0050   $\pm$   0.0082   $\pm$    0.0013\\
0.0080 & 1&6 & 0.9789   $\pm$   0.0031   $\pm$   0.0037
& 0.0012   $\pm$   0.0050    $\pm$   0.0012 \\
0.0125 & 2&2 & 0.9757   $\pm$   0.0029   $\pm$   0.0027
& 0.0033   $\pm$   0.0043    $\pm$   0.0009 \\
0.0175 & 2&9 & 0.9798   $\pm$   0.0032   $\pm$   0.0024
& 0.0002   $\pm$   0.0045    $\pm$   0.0010 \\
0.025 & 3&5 & 0.9723   $\pm$   0.0025   $\pm$   0.0022
& 0.0053   $\pm$   0.0035    $\pm$   0.0009 \\
0.035 & 4&6 & 0.9621   $\pm$   0.0028   $\pm$   0.0019
& 0.0029   $\pm$   0,0037    $\pm$   0.0008 \\
0.050 & 5&8 & 0.9586   $\pm$   0.0023   $\pm$   0.0018
& 0.0050   $\pm$   0.0030    $\pm$   0.0008 \\
0.070 & 7&3 & 0.9449   $\pm$   0.0027   $\pm$   0.0017
& --0.0073   $\pm$   0.0035    $\pm$   0.0008 \\
0.090 & 8&5 & 0.9344   $\pm$   0.0031   $\pm$   0.0017
& --0.0030   $\pm$   0.0039    $\pm$   0.0008 \\
0.110 & 9&6 & 0.9227   $\pm$   0.0035   $\pm$   0.0018
& --0.0003   $\pm$   0.0042    $\pm$   0.0007 \\
0.140 &10&9 & 0.9094   $\pm$   0.0028   $\pm$   0.0018
& --0.0103   $\pm$   0.0033    $\pm$   0.0008 \\
0.180 &12&5 & 0.8814   $\pm$   0.0033   $\pm$   0.0018
& --0.0116   $\pm$   0.0037    $\pm$   0.0010 \\
0.225 &13&8 & 0.8587   $\pm$   0.0035   $\pm$   0.0019
& --0.0054   $\pm$   0.0040    $\pm$   0.0013 \\
0.275 &16&7 & 0.8370   $\pm$   0.0042   $\pm$   0.0022
& --0.0082   $\pm$   0.0049    $\pm$   0.0016 \\
0.350 &19&6 & 0.8015   $\pm$   0.0039   $\pm$   0.0023
& --0.0119   $\pm$   0.0045    $\pm$   0.0020 \\
0.450 &23&4 & 0.7629   $\pm$   0.0057   $\pm$   0.0026
& --0.0099   $\pm$   0.0071    $\pm$   0.0024 \\
0.550 &25&7 & 0.7327   $\pm$   0.0085   $\pm$   0.0029
& --0.0035   $\pm$   0.0115    $\pm$   0.0024 \\
0.675 &26&9 & 0.7202   $\pm$   0.0112   $\pm$   0.0036
& --0.0115   $\pm$   0.0168    $\pm$   0.0031 \\
\hline
\end{tabular}
\end{center}
}
\caption{The ratio $F_{2}^{\rm d}/F_{2}^{\rm p}$ 
and the logarithmic slopes
$b_2=\mbox{d}(F_{2}^{\rm d}/F_{2}^{\rm p})/\mbox{d}\ln Q^2$
determined at $x$ and $\langle Q^2 \rangle$. }
\label{tab:f2npx}
\end{table}
\newpage

\newpage

\newpage

\newpage

\newpage

\newpage

\newpage

\newpage
\end{document}